\documentclass[a4paper,12pt]{book}
\usepackage[a4paper,top=3cm,left=3cm,right=2cm,bottom=2cm]{geometry}

\usepackage{amssymb,amsmath}
\usepackage{bbold}
\usepackage{graphicx}
\usepackage{chngcntr}
\usepackage{tikz}
\usepackage{tikz-3dplot}
\usepackage{indentfirst}
\usepackage[numbers,sort&compress]{natbib}
\usepackage[title]{appendix}
\usepackage{subfig}
\usepackage{floatrow}
\usepackage{booktabs}
\usepackage{mathtools}
\usepackage{blindtext}
\usepackage{imakeidx}
\usepackage{fancyhdr}
\usepackage{lipsum}
\usepackage{chngcntr}
\usepackage{multirow}
\usepackage{setspace}
\usepackage{hyperref}
\usepackage{times}
\usepackage{enumerate}
\usepackage{epigraph}

\numberwithin{figure}{chapter}
\numberwithin{table}{chapter}

\makeatletter
     \renewcommand*\l@figure{\@dottedtocline{1}{1em}{3.2em}}
\makeatother

\makeindex[columns=1]
\usetikzlibrary{trees}
\usetikzlibrary{decorations.pathmorphing}
\usetikzlibrary{decorations.markings}

\tikzset{
    photon/.style={decorate, decoration={snake}, draw=black},
    fermion/.style={draw=black, postaction={decorate}, 
    decoration={markings,mark=at position .55 with {\arrow[scale=2,draw=black]{>}}}},
    afermion/.style={draw=black, postaction={decorate}, 
    decoration={markings,mark=at position .55 with {\arrow[scale=2,draw=black]{<}}}},
    gluon/.style={decorate, draw=green,
    decoration={coil,amplitude=4pt, segment length=5pt}} 
    }

\usepackage{afterpage}

\renewcommand{\and}{\\}
\newcommand{\trento}{$\rm T_{R}ENTo$}
\newcommand{\mychapter}[1]{\chapter{#1}\markboth{#1}{}
}
\newcommand{\mysection}[1]{\section{#1}
}
\newcommand{\mysubsection}[1]{\subsection{#1}
}
\newcommand{\qcd}{Quantum Chromodynamics }
\newcommand{\qgp}{Quark-Gluon Plasma }

\title{Title}
\author{Fabio de Moraes Canedo}

\begin{document}
\pagenumbering{roman}

{
\pagestyle{empty}
\begin{center}

	{\fontsize{16}{16} \selectfont University of S\~ao Paulo \\}
	\vspace{0.1cm}
	{\fontsize{16}{16} \selectfont Physics Institute}
    \vspace{3.3cm}

	{\fontsize{22}{22}\selectfont Study of Jet Quenching in Relativistic Heavy-Ion Collisions
\par}
    \vspace{1cm}


    {\fontsize{18}{18}\selectfont Fabio de Moraes Canedo\par}

    \vspace{1cm}

\end{center}

\leftskip 6cm
\begin{flushright}	
\leftskip 6cm
Supervisor: Prof. Dr. Marcelo G. Munhoz 
\leftskip 6cm
\end{flushright}	

    \vspace{0.2cm}    


\par
\leftskip 6cm
\noindent {Dissertation submitted to the Physics Institute of the University of S\~{a}o Paulo in partial fulfillment of the requirements for the degree of Master of Science.}
\par
\leftskip 0cm
\vskip 0.5cm


\noindent Examining Committee: \\
\noindent Prof. Dr. Marcelo G. Munhoz - Supervisor (IF-USP)\\
Prof. Dr. Sandra S. Padula (IFT-UNESP)\\
Prof. Dr. Tiago Nunes (CFM-UFSC)\\
\vspace{0.5cm}

\centering
    {S\~ao Paulo \\  2020}
\clearpage
}

\chapter*{}

\begin{center}
\textbf{\textit{\`{A} Malu}}
\end{center}

\chapter*{Agradecimentos}

Agrade\c{c}o o meu orientador, Marcelo G. Munhoz, pelo suporte e pelas discuss\~{o}es ao longo desse per\'{i}odo. Agrade\c{c}o tamb\'{e}m pela paci\^{e}ncia com meu estilo ca\'{o}tico de trabalho. Muitas de nossas conversas trouxeram-me grande paz de esp\'{i}rito e conforto para a minha ansiedade. Oxal\'{a} continuemos trabalhando juntos por bastante tempo. Agrade\c{c}o a Jacquelyn Noronha-Hostler e o Jorge Noronha pelas discuss\~{o}es. Agrade\c{c}o também os membros do HEPIC, em especial Cristiane Jahnke, Geovane Grossi, Henrique Zanoli, Lucas Teixeira e Ricardo Pitta. V\'{a}rios bugs n\~{a}o teriam sido resolvidos sem as minhas conversas com voc\^{e}s. Agrade\c{c}o o Ricardo Rom\~{a}o da Silva pelo apoio tamb\'{e}m em v\'{a}rios momentos com o SAMPA.

Agrade\c{c}o aos professores que tive durante a minha vida, que estimularam minha sede por f\'{i}sica e por conhecimento em geral. Nossos encontros fizeram de mim uma pessoa melhor. Em especial os professores Jailton Ara\'{u}jo Oliveira e S\'{e}rgio Souza. Agrade\c{c}o os colegas que tive tamb\'{e}m durante a gradua\c{c}\~{a}o, em especial a Carol Guandalim e o Victor Gomes Da Costa Lobo.

Agrade\c{c}o a minha companheira Malu, pelo carinho e amor. Pelos \'{o}timos momentos que tivemos juntos, pelos conselhos que me deu nessa jornada e por suportar minhas presepadas. Agrade\c{c}o tamb\'{e}m pelo apoio que me deu na revis\~{a}o deste trabalho. Agrade\c{c}o meus pais Vitor e Marcia por tudo. Agrade\c{c}o tamb\'{e}m o suporte e amor incondicional que me deram e continuam me dando. Agrade\c{c}o meus irm\~{a}os Felipe e Mikael por suportarem com gra\c{c}a, na maioria das vezes, minhas brincadeiras. Agrade\c{c}o ao Jacinto, meu salsicho, pelo calor e carinho enquanto redigi esse trabalho. Agrade\c{c}o os meus grandes amigos Gabriel Zoha e Rayner Ribeiro pelas discuss\~{o}es sobre f\'{i}sica, matem\'{a}tica e pela espor\'{a}dica companhia alc\'{o}lica.

Enfim, agrade\c{c}o ao CNPq, pelo aux\'{i}lio financeiro, sob o processo 132927/2018-7, sem o qual essa disserta\c{c}\~{a}o n\~{a}o poderia ter sido realizada.

\chapter*{}

\epigraph{\textit{In so far as a scientific statement speaks about reality, it must be falsifiable; and in so far as it is not falsifiable, it does not speak about reality.}}{Karl Popper}

\epigraph{\textit{``One of the symptoms of an approaching nervous breakdown
			      is the belief that one’s work is terribly important."}
			      }{Bertrand Russell, \textit{The Conquest of Happiness}}

\chapter*{Resumo}

Neste trabalho n\'{o}s investigamos poss\'{i}veis impactos que o plasma de Quarks e Gl\'{u}ons pode ter nos observ\'{a}veis de Jatos. N\'{o}s escolhemos o JEWEL (Jet Evolution With Energy Loss) para este estudo. N\'{o}s acoplamos o JEWEL com o modelo $\rm T_RENTo$ e tamb\'{e}m com o MC-KLN+vUSPhydro para as simula\c{c}\~{o}es. As simulaç\~{o}es foram realizadas para colis\~{o}es chumbo-chumbo a energia $\sqrt{s_{NN}} = 2.76 \, {\rm TeV}$ para centralidade $0-10\%$. Nessas condi\c{c}\~{o}es, observ\'{a}veis de forma e geometria dos jatos n\~{a}o s\~{a}o modificados pela implementa\c{c}\~{a}o de uma hidrodin\^{a}mica e condi\c{c}\~{o}es iniciais realistas. Tamb\'{e}m calculamos o $v_2$ dos jatos. Neste caso n\'{o}s conclu\'{i}mos que as condi\c{c}\~{o}es iniciais tamb\'{e}m n\~{a}o afetam esse observ\'{a}vel. No caso da hidrodin\^{a}mica realista, houve uma melhoria na descri\c{c}\~{a}o desse observ\'{a}vel.

\noindent\textbf{Palavras-chave:} \'{I}ons Pesados; Plasma de Quarks e Gluons; Hidrodin\^{a}mica; Supress\~{a}o de Jatos; Cromodin\^{a}mica Qu\^{a}ntica.

\chapter*{Abstract}

In this work, we investigate possible impacts that the behavior of the Quark-Gluon Plasma might have on Jet Observables. We choose JEWEL (Jet Evolution With Energy Loss) for this study. We have coupled JEWEL with $\rm T_RENTo$ and also with MC-KLN+vUSPhydro for simulations. The simulations were performed for PbPb collisions at $\sqrt{s_{NN}} = 2.76 \, {\rm TeV}$ in the $0-10\%$ centrality class. We have found that jet shape observables are mainly unchanged by the inclusion of realistic hydrodynamics and initial conditions in these settings. We also made calculations for the jet $v_2$. In this case, we have found that initial conditions do not affect this observable. In the case of realistic hydrodynamics, there is an improvement in the description of data. 

\noindent\textbf{Keywords:} Heany-Ion; Quark-Gluon Plasma; Hydrodynamics; Quantum Chromodynamics.

\tableofcontents

\listoffigures



\mychapter{Introduction}

\pagenumbering{arabic}

One of the key problems in particle physics today is the understanding of confinement, property of the fundamental particles that interact through the strong force. This property was originally hypothesized to explain the fact that no particles of fractional charge were ever observed in accelerators and cosmic rays experiments and it was a necessity of the quark model\cite{halzen_quarks_1984}. When the quark model was combined with QCD(\qcd), it became a major puzzle to explain confinement in light of this theory.
\par
Jets are a consequence of confinement. The way they are generated is through the following mechanism. In accelerator experiments, two partons might interact by exchanging a large momentum, a phenomenon known as hard scattering. When this happens and the quarks start to move away, their potential energy starts to grow linearly. This is different from electromagnetism in which the energy falls as $r^{-1}$. Due to this increase in energy, it eventually becomes favorable for the system to excite a pair of quarks from vacuum\cite{dissertori_quantum_2003}. The system evolves into a new pair of color neutral partons, which are closer together. This process occurs iteratively until the resulting pairs don't have enough energy to increase their distance. The result, experimentally, is a spray of particles that hit the detector with similar angles. These are called jets.
\par
The understanding of confinement necessarily requires also the understanding of QCD vacuum properties as was pointed out by T. D. Lee\cite{lee_abnormal_1975}:

\begin{quote}
To study the question of ``vacuum", we must turn to a different direction, we should investigate some ``bulk" phenomena by distributing high energy over a relatively large volume.
\end{quote}

The excited state of matter, in which quarks and gluons are not confined, is called Quark-Gluon Plasma. This state of matter can be created in the laboratory today through ultra-relativistic heavy-ion collisions. The idea is that during the collision, a good part of the kinetic energy is converted into thermal energy, and a phase transition occurs. Later on, the matter expands and cools down, transitioning back into hadrons, which are bound states of quarks and gluons. These final hadrons eventually stop interacting as well and start their travel as free particles. They potentially go through further decays of electromagnetic or weak nature and then hit the detectors.

\begin{figure}
\includegraphics[width=1\textwidth]{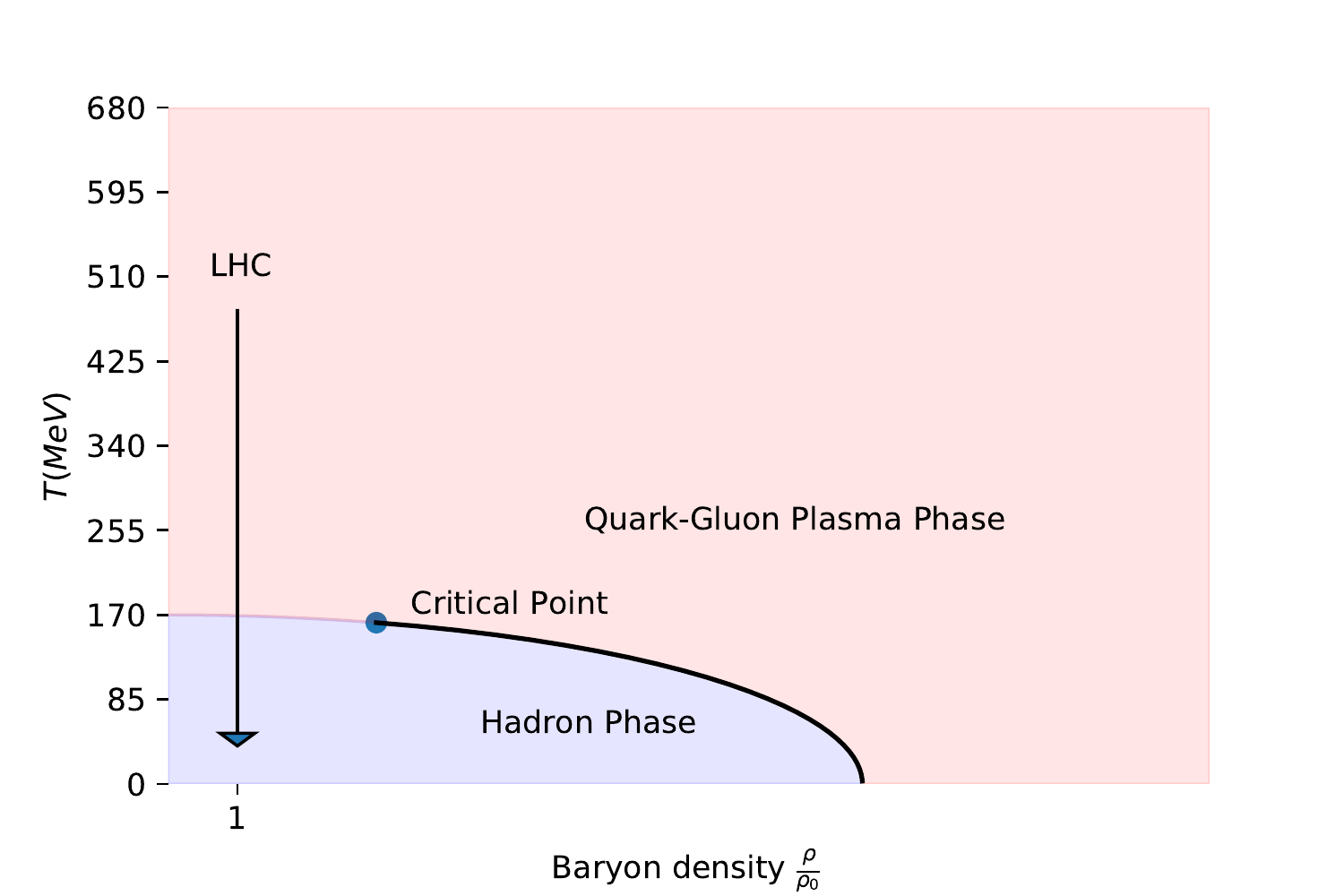}
\caption{QCD phase diagram}
\label{phase_diag}
\end{figure}

Once the particles hit the detector, they might be identified and their properties measured, such as their transverse momentum and their rapidity. The distribution of these particles is then analyzed to try to infer what has happened during the collision. One phenomenon that was identified in the study of this final state is the so-called Jet Quenching. It corresponds to the effects of the medium present in heavy-ion collisions on the hard scattering. When hard scattering occurs in heavy-ion collisions, the partons are surrounded by a color excited medium. Like in an ordinary plasma, a screening effect prevents their potential energy to grow linearly. They interact further with the hot and dense medium created in their surroundings before experiencing the process of jet creation. In this interaction, they lose energy due to elastic scattering and \emph{gluonstrahlung}, resulting in jets with less energy and broader.
\par
In this work, a study was performed of observables related to Jet Quenching that could be sensitive to the finer details of the initial conditions as well as the hydrodynamic evolution of the quark-gluon plasma formed in ultra-relativistic heavy-ion collisions. Several known observables were analyzed, both from the perspective of the inner jet substructure and shape, as well as soft-high $p_T$ correlation observables such as the jet $v_2$. Since that involves collective behavior associated with the soft sector.
\par
In Chapter \ref{theory}, the theory and the main models used in this work are explained. In Chapter \ref{method}, the observables, as well as the techniques used to extract and analyze them are explained. In Chapter \ref{results}, the results of the work are presented. In Chapter \ref{conclusions}, the conclusions and discussion of the results are presented.

\mychapter{Theory} \label{theory}

\mysection{\qcd}

Out of the four known interactions given by the current picture of physics, the one that is of the most importance to this work, is the Strong Force\index{Strong Force}. This force is the one responsible for the nuclear stability. It is described by the following lagrangean density\citep{peskin_introduction_1995,halzen_quarks_1984}:

\begin{equation}
\mathcal{L}_{\rm QCD} = -\frac{1}{4}F^{A}_{\alpha\beta}F^{\alpha\beta}_{A}
+\sum_{flavors}
\bar{q}_{A}(i\gamma^{\mu}D_{\mu}-m)_{AB}
q_{B}
\end{equation}

where $F^{\alpha\beta}_{A}$ is the gluon field tensor, $q_{B}$ and $\bar{q}_{A}$ are the fermion fields and $D_{\mu}=\partial_{\mu} + i A_{\mu}/g$ is the covariant derivative. It describes six fermion fields, corresponding to the known fundamental quarks, and also describes their coupling with a non-abelian Yang-Mills field with the \textbf{SU(3)} symmetry. This theory, given its non-abelian nature, has interesting properties that do not appear in electromagnetism.
\par
At first, it is a theory with asymptotic freedom\index{asymptotic freedom}, i.e., the coupling constant\index{coupling constant} decreases as the process under consideration has a larger momentum transfer. This allows one to apply the formalism of perturbation theory\index{perturbation theory} in the calculation of high energy processes. Also, it has non-linear terms in the Yang-Mills sector due to the non-commutative nature of the group. These terms describe the field self-interaction, and also make the theory particularly troublesome to allow calculations. They are also responsible for the decrease of the coupling constant aforementioned. As a result, for low energy processes, perturbation theory is no longer reliable. Effective models, semi-analytical and numerical methods, such as lattice QCD, are then used to try to make predictions. These are classified as non-perturbative methods.
\par
As already mentioned, the coupling constant of the theory varies with energy. The behavior of the coupling constant as one varies the energy of a given process was calculated and is given by\cite{dissertori_quantum_2003}:

\begin{equation}
\alpha_s (q^2) = \frac{2}{b_0 \ln (-q^2/\Lambda^2)}
\end{equation}

The sign of $b_0$ in this equation is opposite to that of QED. As a result, the strength of the interaction grows with distance and is unbounded. The scale $\Lambda$ is a generated scale of QCD and does not depend on quark masses. It is believed that this dynamical scale alone is responsible for the hadron masses, and not the quark masses.

\mysection{\qgp} \label{qgp}

The work by E. Fermi\cite{fermi_high-energy_1950} and L. Landau\cite{lifshitz_statistical_1980,haar_collected_2013} on multiparticle production have paved the way for Hagedorn to develop his bootstrap model\index{model!Hagedorn} in the early sixties. There was already an idea to describe particle physics through the formalism of hydrodynamics. For this to be possible, an understanding of the mass spectrum\index{mass spectrum} of hadrons was necessary. Hagedorn realized that it was possible to find such a spectrum by understanding that heavy hadrons are bound states of lighter hadrons(which are bound states of lighter hadrons(which are bound states of lighter hadrons(...))). With this idea, he developed a formalism that arrived at the following expression for the mass spectrum:

\begin{equation}
\rho(m) \approx c(m_0^2 + m^2)^{k/2} \exp(m/T_0)
\end{equation}

The parameters were meant to be fixed by experiments. The important consequence of this mass spectrum was that the thermodynamical potentials of a hadron gas would have a singularity as $T \rightarrow T_0$. This was the first indication of a phase transition.

\par
In the eighties\cite{letessier_hadrons_2002}, with the development of the theoretic understanding of \qcd, it was predicted that, at a certain temperature, hadrons would ``melt" and partons would no longer be confined. This would then induce a phase transition, bringing about a new state of matter, in which quarks and gluons would be the fundamental degrees of freedom of the partition function. Experimental efforts were put forth to create this new state in the laboratory and, by 2005, RHIC had gathered conclusive data that showed this new state of matter had been created in its experiments\cite{Adams:2005dq}.
\par
A few signatures of the \qgp, name assigned to this new state, were found in these experiments\cite{letessier_hadrons_2002}. First, there was the so-called azimuthal anisotropic flow\index{anysotropic flow}. A phenomenon theoretically explained as the conversion of the asymmetry of the geometric distribution of energy in the transverse plane into an asymmetry in momentum space. Besides, there was the strangeness enhancement\index{strangeness} of the observed particles, a hint that a possible chemical equilibrium had been achieved during the collisions. There was also the phenomenon of Jet Quenching, commonly explained by the fact that, unlike in proton collisions, the partons generated in the hard scattering must traverse a dense and hot medium. Its main signature was the suppression of high transverse momentum particles.
\par
\begin{figure}
\includegraphics[width=0.5\textwidth]{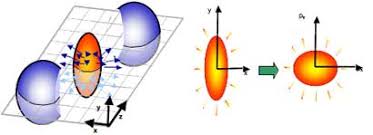}
\caption[Azimuthal anizotropy representation]{Azimuthal anizotropy representation, picture from \cite{ferrer_exploring_2016}}
\label{v2}
\end{figure}

The azimuthal anisotropic flow, as previously explained, is related to the asymmetry of the collision. It can be better understood by observing Figure \ref{v2}. In the Figure, it is shown that in the interaction region, in orange, there is a preferential direction for the energy deposition. This geometric property of the distribution is called ellipticity. It is responsible for a larger pressure gradient in the $x$ direction. The gradient is the force responsible for the conversion of the asymmetry from the position space into momentum space. The intensity of this conversion can be calculated through hydrodynamic models. This allows the extraction of properties of the \qgp, such as its viscosity and EOS(Equation of State).

The strangeness enhancement\index{strangeness!enhancement} is explained through the chemical equilibrium that is achieved in heavy-ion collisions\cite{letessier_hadrons_2002}. This means that the strange particles are generated through thermodynamical processes, not just in the hard-scattering in the early times of the collision. Also, they are not particles that come from the sea quarks inside the colliding nucleons. Normally, this enhancement is measured by a quantity called $ \rm R_{\rm AA} $\index{$ \rm R_{\rm AA} $}, that is defined as the ratio of transverse momentum from pp(proton-proton) to AA(nucleus-nucleus) collisions:

\begin{equation}
{\rm R_{AA}}(p_T,\eta) = \frac{\frac{dN_{AA}}{d{p_T}d\eta}}{\langle N_{bin}\rangle \frac{dN_{pp}}{d{p_T}d\eta}}
\end{equation}

Where $N_{AA}$ is the number of counts (in this case, of strange particles) in AA collisions, $N_{pp}$ is the same count on pp collisions, and $\langle N_{bin}\rangle$ is the average number of binary collisions that happen on a nucleus-nucleus collision, usually estimated with Glauber model.
\par
\begin{figure}
\includegraphics[width=0.5\textwidth]{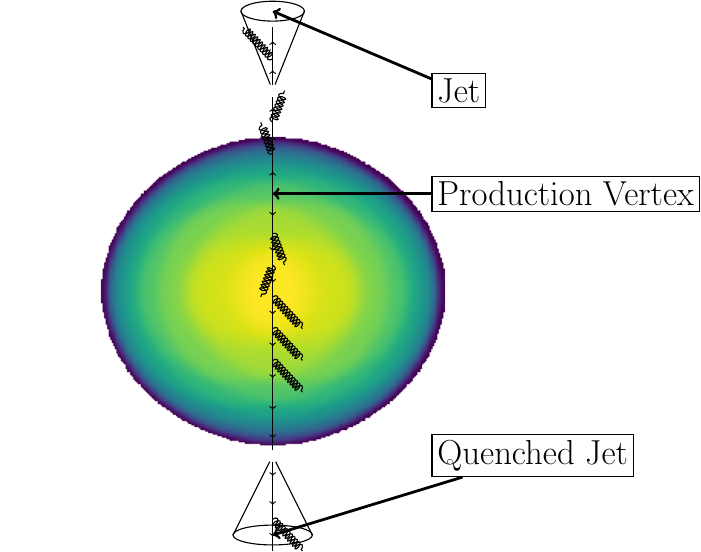}
\caption[Schematics of a parton traversing the medium.]{Schematics of a parton traversing the medium.}
\label{arrefecimento}
\end{figure}

The Jet Quenching phenomena\index{Jet!Quenching} is observed through the $\rm R_{AA}$ taken from the $p_T$ spectrum of particles. This is theoretically interpreted through the parton energy loss while they traverse the plasma, as illustrated in Figure \ref{arrefecimento}. Beyond the spectrum suppression, jet analysis has also revealed that there are distinct fragmentation patterns, as opposed to the vacuum case, hinting a parton-medium interaction\cite{denterria_jet_2009}.

The description of the medium that JEWEL (Jet Evolution With Energy Loss) uses, in its default settings, consists of calculating the partition function by assuming an ideal quantum free gas of non-interacting particles. This model is based on the idea of asymptotic freedom\index{asymptotic freedom}, which is the decrease of the interaction of the partons with the increase of temperature.
\par
This model is necessary to translate temperature profiles into a scattering center density. This is done though the following equation\cite{bjorken_highly_1983,zapp_jet_2006}:

\begin{equation}
n_s= 2 g_g \zeta(3) \frac{T^3}{2 \pi^2}
\end{equation}

Results calculated with the lattice \qcd approach corroborate that this expressions are valid for $T = 200 \, \rm MeV$\cite{letessier_hadrons_2002}. Before that, at $T \approx 185 \, \rm MeV$, there is a rapid transition from this fluid to the free hadron gas\index{critical temperature}. These equations for an ideal relativistic quantum gas are used in some Jet Quenching models\footnote{such as JEWEL(Jet Evolution With Energy Loss)} for the description of the medium.

\mysubsection{Hydrodynamics of the \qgp}

Hydrodynamics is the formalism chosen to model the evolution from thermalization until the free streaming of particles. The conversion of hydrodynamic calculations into free hadrons is made with the Cooper-Frye prescription\index{prescription!Cooper-Frye}. Hydrodynamics is essentially approached through the relativistic energy-momentum tensor\cite{florkowski_phenomenology_2010}

\begin{equation}
T^{\mu \nu} = \epsilon u^{\mu} u^{\nu} + P \Delta^{\mu \nu}
\label{tensor}
\end{equation}

Where $\Delta^{\mu \nu} = g^{\mu \nu} - u^\mu u^\nu$ is the projection operator orthogonal to $u^\nu$, $P$ is the local pressure and $\epsilon$ is the energy density. The conservation equation\index{conservation equation} for this tensor is:

\begin{equation}
\partial_{\mu} T^{\mu \nu} + \Gamma_{\mu \lambda}^{\nu} T^{\mu \lambda} = 0
\label{conservation_eq}
\end{equation}

This equation has, in total, six independent quantities, four from the local velocity plus two from energy density and pressure. There are four equations from energy-momentum conservation, which leaves two unknowns. The normalization of the four-velocity plus the equation of state then solve the system.
\par
When the system is out of equilibrium, the energy-momentum tensor is different from the one in equation \eqref{tensor}. In the first order, it assumes the form:

\begin{equation}
T^{\mu \nu} = \epsilon u^{\mu} u^{\nu} + P \Delta^{\mu \nu} + \tau^{\mu \nu}
\end{equation}

Where $\tau^{\mu \nu}$ corresponds to a deviation from equilibrium state. This form is constructed by creating an artificial equilibrium state by considering the local energy density, which is a well-defined quantity\cite{romatschke_new_2010}. Here $\tau^{\mu \nu}$ is required to be symmetric due to the conservation of angular momentum. Also, it is required to satisfy $\tau^{\mu \nu} u_{\mu} =0$, according to Laudau definition of the four-velocity. Given these properties, $\tau^{\mu \nu}$ can be written as:

\begin{equation}
\tau^{\mu \nu} = \pi^{\mu \nu} + \Pi \Delta^{\mu \nu}
\end{equation}

At this point, it is useful to define the projection operator:

\begin{equation}
\Delta^{\alpha \beta}_{\mu \nu} = \frac{1}{2} \left[ \Delta^{\alpha}_{\mu}\Delta^{\beta}_{\nu} + \Delta^{\alpha}_{\nu}\Delta^{\beta}_{\mu} - \frac{2}{3} \Delta^{\alpha \beta}\Delta_{\mu \nu} \right]
\label{projection}
\end{equation}

It is symmetric in $\alpha$ and $\beta$ and in $\mu$ and $\nu$, and the resulting projection is traceless. This operator helps us to decompose $\tau^{\mu \nu}$ in two different components with simple physical interpretations.

\begin{subequations}
\begin{equation}
\tau^{\alpha \beta} = \Delta_{\mu \nu}^{\alpha \beta} \pi^{\mu \nu}
\end{equation}

\begin{equation}
\Pi \Delta^{\alpha \beta} = \pi^{\alpha \beta} - \tau^{\alpha \beta}
\end{equation}
\end{subequations}

The first one corresponds to vorticity in the flow, and the second one corresponds to the expansion and compression of the fluid.
\par
The code to integrate numerically the equations \eqref{conservation_eq}. It will have to be provided with differential equations for the extra degrees of freedom included in $\tau^{\mu \nu}$. This part is usually chosen according to a specific model. The equations used in this work are:

\begin{subequations}
\begin{equation}
\tau_{\Pi}(D\Pi + \Pi\theta) + \Pi + \zeta\theta=0
\label{bulk_visc}
\end{equation} 

\begin{equation}
\tau_{\tau}(\Delta_{\mu \nu \alpha \beta}\tau^{\alpha \beta} + \frac{4}{3}\tau_{\mu \nu}\theta) + \tau^{\mu \nu} = 2\eta \Delta_{\mu \nu \alpha \beta}\Delta^{\alpha} u^{\beta}
\label{shear_visc}
\end{equation} 
\end{subequations}

These are the simplest equations that are both causal and stable.
The parameters $\tau_{\Pi}$ and $\tau_{\tau}$ are lifetimes for the out-of-equilibrium terms. $\zeta$ if the bulk viscosity coefficient, and $\eta$ is the shear viscosity coefficient. $\theta$ is the four-divergence of the four-velocity, and $D=u^{\mu}D_{\mu}$, where $D_{\mu}$ is the covariant derivative, which, for scalar functions, as is the case in Equation \eqref{bulk_visc}, is equal to $\partial_{\mu}$.
\par
Besides the conservation equation \eqref{conservation_eq}, there might be other currents (Such as baryon number, strangeness number, etc) that satisfy a conservation equation as well:

\begin{equation}
\partial_{\mu} j^{\mu} + \Gamma_{\mu \nu}^{\nu} j^{\mu} = 0
\end{equation}

Each current included also comes with a conservation equation, so the system is still solvable.

\mysection{Relativistic Heavy-Ion Collisions}

\begin{figure}
\includegraphics[width=1.0\textwidth]{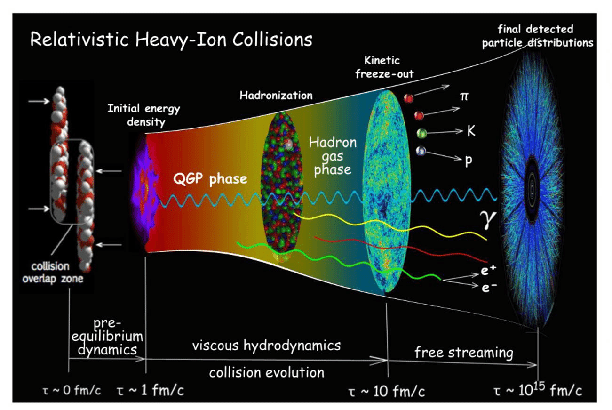}
\caption[Heavy-ion collision scheme]{Heavy-ion collision scheme, picture from \cite{scharenberg_hot_2018}}
\label{heavy_ion}
\end{figure}

The experimental method used to study the QGP is the collision of heavy nuclei at relativistic velocities. The general picture for such a collision can be seen in Figure \ref{heavy_ion}. Two Lorentz contracted nuclei approach each other at relativistic speed and then collide, leaving energy in the interaction region. The matter deposited in the interaction region then thermalizes in about $\sim 1 {\rm fm/c}$. This stage is currently the most poorly understood and the source of the largest uncertainty. After thermalization, the hot system expands until $\sim 10 {\rm fm/c}$, a stage that can be simulated with hydrodynamics codes using as input the models that attempt to describe the initial energy density profile. During the hydro evolution, the system reaches the hadronization stage. After the hadron stage the kinetic freeze-out happens in which particles start to stream freely until they reach the detectors at about $\sim 10^{15} {\rm fm/c}$.

\mysubsection{Initial Conditions} \label{IC}

For the initial conditions, as explained, models are used to predict the energy density profiles that result from the early dynamics of the collision. One of the first models used was Glauber wounded nucleon model\cite{miller_glauber_2007}. Later, models that implemented sophisticated ideas inspired by QCD were developed. Here we studied three different models. First, we used JEWEL\cite{zapp_monte_2009,zapp_local_2011} default model, which is a Glauber like without any type of fluctuation in the nucleons positions. Then, we used \trento\cite{moreland_alternative_2015}, which is also Glauber inspired, but parametric in nature, and can be tuned to fit certain experimental results and extract qualitative behavior of the initial conditions. And then we used MC-KLN\cite{drescher_effects_2007}, based on the color-glass condensate formalism.

\begin{figure}
\subfloat[Glauber]{
\includegraphics[width=0.3\textwidth]{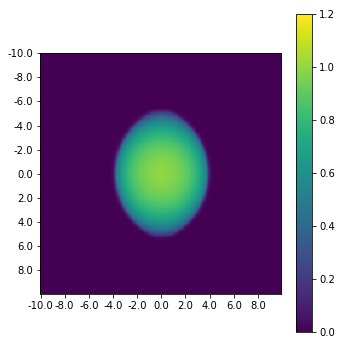}\label{ic_glauber}
}
\label{glauber}
\subfloat[MC-KLN]{
\includegraphics[width=0.3\textwidth]{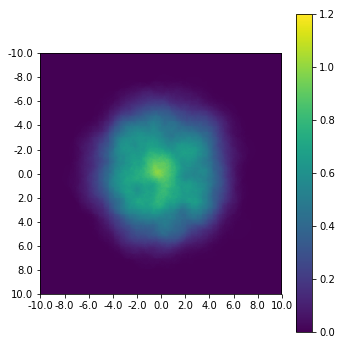}\label{ic_kln}
}
\subfloat[\trento]{
\includegraphics[width=0.3\textwidth]{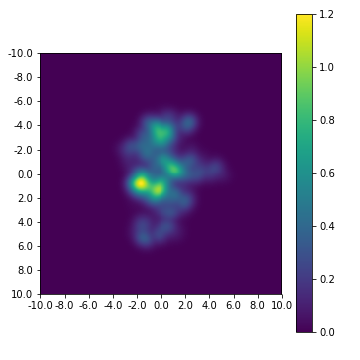}\label{ic_trento}
}
\caption{Energy density in arbitrary units for different initial condition models. The energy density is displayed in arbitrary units.}
\label{ic}
\end{figure}

\mysubsection{Glauber smooth model} \label{glauber}

The implementation of the background medium used by JEWEL is an idealized and smooth Glauber model. For nuclei density, a Wood-Saxon distribution is assumed:

\begin{equation}
\rho({\overrightarrow{\bf \it r}}) = \frac{1}{1+\exp(\frac{r-R}{a})}
\end{equation}

Where $R$ is the nuclei radius and $a$ represents a \emph{skin-depth} of the nuclei. This profile density if almost entirely constant for $r<R$. The transverse density is taken from the longitudinal integration:

\begin{equation}
T_A(x,y) = \int dz \, \frac{1}{1+\exp(\frac{r(x,y,z)-R}{a})}
\end{equation}

This $T_A$ is called the thickness function. The combination of the thickness function of both nuclei then determines the participant density:

\begin{equation}
\begin{split}
n (b,x,y) &\propto T_A(x-\frac{b}{2},y) \left[ 1 - \exp \left( - \sigma_{NN} T_B(x+\frac{b}{2},y) \right) \right] \\ & + T_B(x+\frac{b}{2},y) \left[ 1 - \exp \left( - \sigma_{NN} T_A(x-\frac{b}{2},y) \right) \right]
\end{split}
\end{equation}

Where $\sigma_{NN}$ is the nucleon-nucleon cross-section. An example of this participant profile can be observed in Figure \ref{ic}(a). The name smooth used here references the lack of \emph{lumpiness} in this profile, as compared to the others in Figure \ref{ic}.

\mysubsection{T$_{\rm R}$ENTo} \label{trento}

T$_{\rm R}$ENTo\cite{moreland_alternative_2015} is a model for initial conditions for heavy-ion collisions based on the Glauber model. It is based on the idea of the thickness function:

\begin{equation}
T_A (x,y) = \int {\rm d}z \, \rho^{\rm part} (x,y,z)
\end{equation}

Which is built from the density of participants $\rho^{\rm part} (x,y,z)$. Once the thickness functions of both nuclei are defined, a scalar function is defined as:

\begin{equation}
f = \left( \frac{T^p_A+T^p_B}{2} \right)^{1/p}
\end{equation}

\begin{figure}
\includegraphics[width=1.0\textwidth]{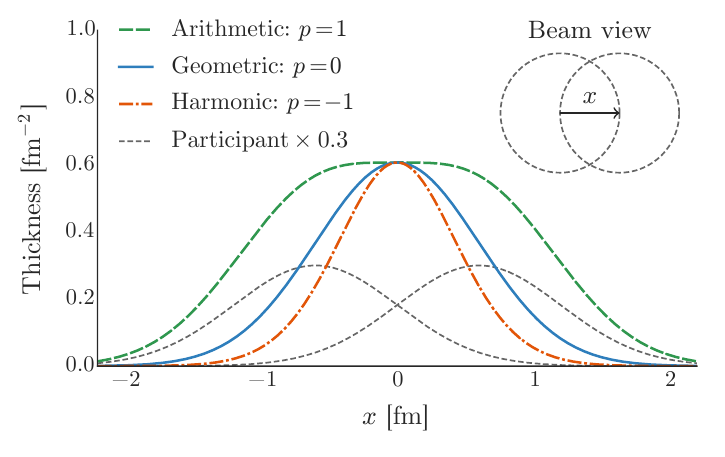}
\caption[$T_RENTo$ reduction thickness.]{$T_RENTo$ reduction thickness. The dashed lines correspond to nucleon profiles and the colored lines correspond to the reduced thickness for different values of $p$. Picture from \cite{moreland_alternative_2015}}
\label{trento_thick}
\end{figure}

Where $p$ is a free parameter. This is a generalized mean and is taken to be proportional to the entropy deposition. This function has a general behavior that changes continuously with the parameter $p$, interpolating between maximum and minimum. Each value corresponds to a qualitatively different entropy deposit mechanism. The qualitative behavior of this function can be seen in Figure \ref{trento_thick}.

The participant densities are built by considering pairs of nucleons, one from each nucleus. Their positions are sampled from a Woods-Saxon potential. Then for each pair the probability of collision is taken as:

\begin{equation}
P_{coll} = 1 - \exp \left[ -\sigma_{gg} \int dx \, dy \int dz \, \rho_A \int dz \, \rho_B \right]
\end{equation}

Here we have that $\sigma_{gg}$ is adjusted so that the total proton cross-section is $\sigma_{NN}$\cite{moreland_alternative_2015}. If the nucleons do collide, then the thickness function is built from the sum of each nucleon density, which is taken to be a gaussian. So each nucleon contributes with a thickness function given by:

\begin{equation}
T_{N} = w \int dz \, \rho_N (x,y,z) 
\end{equation}

Where $w$ is a random weight sampled from a gamma distribution with unity mean:

\begin{equation}
P_k(w)=\frac{k^k}{\Gamma(k)} w^{k-1} e^{-kw}
\end{equation}

This creates an extra degree of fluctuation that will be encoded in a parameter $k$ that can be chosen to fit experimental results. An example of a $\rm T_RENTo$ profile for $p=0$, which is a tune that reproduces IP-Glasma behavior, can be seen in Figure \ref{ic}(c). The effects of fluctuations can be observed in the figure.

\mysubsection{MC-KLN} \label{mckln}

\begin{figure}
\includegraphics[width=0.8\textwidth]{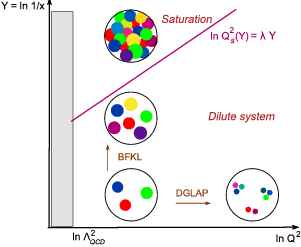}
\caption[Hadron description and its evolution equations.]{Hadron description and its evolution equations.}
\label{cgc}
\end{figure}

The MC-KLN\cite{drescher_effects_2007} model is based on the color glass condensate approach. The key idea is that a nuclei is essentially a gluon wall with a wavelength bigger than the contracted longitudinal size of the other nuclei. So the collision occurs in a coherent manner. This comes from the fact that a hadron can be seen as a collection of partons and the picture we see varies with $x$ and $Q^2$. The description of a hadron with varying these quantities can be seen in Figure \ref{cgc}. The quantity $x$ represents the fraction of the hadron energy carried by the partons. The smaller this fraction, then higher is the number of partons. $Q^2$ can be understood as the transverse momentum carried by each parton, which is negatively correlated with the transverse size of each parton, according to the uncertainty principle. Evolving in $x$ implies then considering the emission of partons by partons. increasing the number of constituents of a hadron. Eventually, though, the cross-section for partons to absorb partons becomes large, and the number of gluons no longer increases.

A saturation scale can be defined such that below it the occupation number of gluons is constant, and above, we have a distribution that obeys the BFKL equations\cite{kovchegov_quantum_2012}. The saturation scale is defined in terms of the thickness function, according to:

\begin{equation}
Q_{s,A}^{2} (x, \textbf{r}_{\perp}) = 2 \, {\rm Gev}^2 \left( \frac{T_A (\textbf{r}_{\perp}) / p_A(\textbf{r}_{\perp}) }{1.53} \right) \left( \frac{0.01}{x} \right)^{\lambda}
\end{equation}

Where $T_A (\textbf{r}_{\perp})$ is the thickness function and $x$ is the fraction of the momentum carried by the gluon in the distribution. $p_A({\mathbf{r}_{\perp}})$ is the probability of finding at least nucleon at a given transverse coordinate. In this way, the model proceeds by randomly picking nucleon positions according to a Woods-Saxon potential, and then it determines the thickness function, finally determining the saturation scales. The gluon distributions are then combined through:

\begin{equation}
\frac{dN_g}{dyd^2\textbf{r}_{\perp}} \sim \int \frac{d^2p_{\perp}}{p^2_{\perp}} \int d^2 k_{\perp} \phi (n_{part,A}) \phi (n_{part,B})
\label{gluon_dist}
\end{equation}

where:

\begin{equation}
\phi(x,k_{\perp}\;\textbf{r}_{\perp}) \sim \frac{1}{\alpha(Q_s^2)} \frac{Q_s^2}{\max (Q_s^2,k_{\perp}^2)}
\label{gdf_kln}
\end{equation}

And $\phi$ is the gluon distribution function. An example of an MC-KLN generated profile can be seen in Figure \ref{ic}(b). The larger spread of the energy density is a consequence of the tail of the gluon distribution in Equation \eqref{gdf_kln}. The ansatz in Equation \eqref{gluon_dist} comes from considering a number of binary collisions between the gluons and a given position in transverse space.

\mysection{Hydrodynamics} \label{hydro}

The implementation of the hydrodynamics involves developing a numerical method for integration of the hydrodynamic equations, as well as an Equation of State and the differential equations satisfied by the independent components of the $\tau^{\mu \nu}$ tensor. Here we used two ways to model this stage of the collision. First, as the JEWEL default, we used the Bjorken longitudinal expansion model, which is only an approximation. For a more realistic treatment, we used the 2+1 code v-USPhydro.

\mysubsection{Bjorken expansion} \label{bjorken}

One of the first models developed to calculate the evolution of the QGP was the one developed by Bjorken that approximates the system to a one-dimensional system and considers the expansion only in the longitudinal direction. This model then predicts a decay of the energy density that can explain the plateau in the rapidity spectrum. The idea is that we have a blast wave in the longitudinal direction that is bound by the speed of sound in the medium:

\begin{equation}
T = T_0 \left( \frac{\tau_0}{\tau} \right)^{v_s^2}
\end{equation}

where $\tau_0$ and $T_0$ are chosen to fit experimental data, mainly the $\frac{dN}{d\eta}$ observable. Since the speed of sound is calculated, for a relativistic ideal quantum gas, to be $v_s^2=\frac{1}{3}$, we have a complete description of the temperature evolution of the system. The values chosen in JEWEL\cite{zapp_perturbative_2013} are $\tau_0=0.5 \, {\rm fm/c}$ and $T_0=530 \, {\rm MeV}$.

\mysubsection{v-USPhydro} \label{vusp}

The hydrodynamics model used here is v-USPhydro\cite{noronha-hostler_bulk_2014,noronha-hostler_bulk_2013,noauthor_jacquelyn_nodate}. It utilizes the Lagrangian method to implement the integration of the conservation equations that describes the hydrodynamics. This method consists, as an alternative to the grid methods, of discretizing the fluid density profile into particles and allowing their positions to evolve as well. One of the advantages of this approach is that it can be efficiently applied to unbound systems, as is the case of nuclei collisions. This formalism is called Smoothed Particle Hydrodynamics.
\par
The idea is to use a finite number of cells to describe the fluid. One then defines a normalized kernel:

\begin{equation}
\int W [\mathbf{\rm r};h]d^2 \mathbf{\rm r} = 1
\end{equation}

where $\textbf{\rm r}$ is a transverse position coordinate and $h$ is a parameter that plays a similar role to the grid spacing in the Euler method. This kernel can then use the cells to calculate the value of a field at any point in space:

\begin{equation}
\tau \gamma \sigma \rightarrow \sigma^* (\mathbf{\rm r}, \tau) = \sum_{\alpha=1}^{N_{SPH}} \nu_\alpha W [ \mathbf{\rm r} - \mathbf{\rm r}_\alpha(\tau);h]
\end{equation}

Where $\tau$ is the proper time, $\gamma$ is the relativistic factor. $\sigma$ is the entropy density and $N_{SPH}$ is the number of Smoothed Particle cells. The conservation of $\tau \gamma \sigma$ is then equivalent to:

\begin{equation}
\sum_{\alpha=1}^{N_{SPH}} \nu_\alpha = K
\end{equation}

where $K$ is a constant and the $\nu_\alpha$ are chosen to match an initial distribution. The entities that occupy the positions $\boldsymbol{\rm r}_{\alpha}$ are referred to as SPH particles. Now, given some extensive quantity associated to the density $a(\boldsymbol{\rm r},\tau)$, the discretized version of it will be:

\begin{equation}
a(\boldsymbol{\rm r},\tau) = \sum_{\alpha=1}^{N_{SPH}} \nu_{\alpha} \frac{a(\boldsymbol{\rm r}_{\alpha} (\tau))}{\sigma^*(\boldsymbol{\rm r}_{\alpha} (\tau))} W [ \boldsymbol{\rm r} - \boldsymbol{\rm r}_\alpha(\tau);h]
\end{equation}

And the derivative of it will be:

\begin{equation}
\frac{d}{d\boldsymbol{\rm r}} a(\mathbf{\rm r},\tau) = \sum_{\alpha=1}^{N_{SPH}} \nu_{\alpha} \frac{a(\boldsymbol{\rm r}_\alpha (\tau))}{\sigma^*(\boldsymbol{\rm r}_\alpha (\tau))} \frac{d}{d\boldsymbol{\rm r}} W [ \boldsymbol{\rm r} - \boldsymbol{\rm r}_\alpha(\tau);h]
\end{equation}

So, a knowledge of the derivative of the kernel function allows one to calculate the derivative of any extensive quantity. Similarly, the proper-time derivative:

\begin{equation}
\frac{d}{d \tau} a(\mathbf{\rm r},\tau) = \sum_{\alpha=1}^{N_{SPH}} \nu_{\alpha} \frac{a(\boldsymbol{\rm r}_\alpha (\tau))}{\sigma^*(\boldsymbol{\rm r}_\alpha (\tau))} \frac{d\boldsymbol{\rm r}_\alpha (\tau)}{d \tau} \frac{d}{d\boldsymbol{\rm r}_\alpha (\tau)} W [ \boldsymbol{\rm r} - \boldsymbol{\rm r}_\alpha(\tau);h]
\end{equation}

The equations of motion for the fluid are\cite{denicol_effect_2009,denicol_effect_2010}:

\begin{subequations}

\begin{equation}
\gamma \frac{d}{d\tau} \left[ \frac{(\epsilon + p + \Pi)}{\sigma} u^{\mu} \right] = \frac{1}{\sigma} \partial^{\mu} (p+\Pi)
\end{equation}

\begin{equation}
\gamma \frac{d}{d \tau} \left( \frac{s}{\sigma} \right) + \left( \frac{\Pi}{\sigma} \right) \frac{\theta}{T} = 0
\end{equation}

\begin{equation}
\tau_{\Pi} \gamma \frac{d}{d\tau}\left( \frac{\Pi}{\sigma} \right) + \frac{\Pi}{\sigma} + \left( \frac{\zeta}{\sigma} \right) \theta = 0
\end{equation}

\begin{equation}
\tau_{\tau}\gamma \Delta_{\mu \nu \alpha \beta} \frac{d}{d\tau} \left( \frac{\tau^{\alpha \beta}}{\sigma} \right) + \frac{\tau_{\mu \nu}}{\sigma} = 2\frac{\eta}{\sigma} \Delta_{\mu \nu \alpha \beta} \bigtriangledown^{\alpha} u^{\beta}
\end{equation}

\end{subequations}

where $\zeta$ and $\eta$ are the bulk and shear viscosity, $\epsilon$, $s$ and $\sigma$ are the energy density, the entropy density, and the system density, respectively. $T$ is the temperature and $\theta$ is the four divergence of the velocity that describes the expansion rate of the system. The EOS used by v-USPhydro is S95n-v1. The coefficients used for the viscosities are $\frac{\eta}{s}=0.3$ and $\frac{\zeta}{s} = 0.08$. v-USPhydro integrates these equations using the discretization procedure described above using a Runge-Kutta method and then calculates freeze-out surface and temperature profile and other quantities. For this work, we are interested in the temperature profile. 

\mysection{Jet Quenching}

\mysubsection{Jet Evolution With Energy Loss (JEWEL)}

JEWEL\cite{zapp_jewel_2014, zapp_monte_2009, zapp_perturbative_2013} is a Monte-Carlo implementation of a perturbative formalism, called the BDMPS-Zakharov formalism which treats elastic collisions and radiation for a parton moving in a dense environment. One of the issues that appears when one wants to consider Monte-Carlo implementations of quantum phenomena is that one must deal with amplitudes, not probabilities. An important consequence of this is the appearance of the so-called QCD LPM (Landau-Pomeranchuk-Migdal) effect. So JEWEL addresses this by the definition of a parameter that will separate the cases where interference might occur. This parameter is called the gluon formation time:

\begin{equation}
\tau = \frac{\mathbf{k}^2}{2\omega}
\end{equation}

Where $\mathbf{k}$ is the radiation momentum, and $\omega$ is the radiation energy. This parameter is used to determine if two successive scatterings are within the formation time of an emission. If that is the case, JEWEL treats the scattering centers as coherent sources of the \emph{gluonstrahlung}.
\par
The algorithm implemented in JEWEL is the following. First, a pair of partons is generated through hard scattering in a point in transverse space proportional to the energy-density. This first hard scattering is generated with PYTHIA\cite{sjostrand_pythia_2006}. Then these partons are propagated through the medium in the transverse plane. In this process they will lose energy through collisions and radiation. After they escape the medium, the event is given back to PYTHIA\cite{sjostrand_pythia_2006} to perform fragmentation and hadronization.
\par
The parton shower is implemented in JEWEL through the virtuality ordering procedures that resums leading logarithm of pQCD. It is implemented through the so-called Sudakov form factor:

\begin{equation}
\mathcal{S}(t_h,t_c) = \exp \left\{ -\int_{t_c}^{t_h} \frac{{\rm d} t}{t} \int_{z_{\rm min}}^{z_{\rm max}} {\rm d} z \sum_b \frac{\alpha_s (k_{\bot}^2)}{2\pi} \hat{P}_{ba} (z) \right\}
\end{equation}

This can be interpreted as the probability that a parton emits no radiation between the scales $t_h$ and $t_c$. The function $\hat{P}(z)$ is the so-called Altarelli-Parisi splitting function. This Sudakov form-factor is used to determine the scale of the next splitting in the parton shower. This shower will behave as if it was in vaccum. It is the collisional process that allows for the generation of further virtuality evolution and will be responsible for medium-modifications of the jet.
\par
The collisional part of JEWEL is treated through simple $2 \rightarrow 2$ matrix elements squared. A thermal mass is attributed to the gluons representing the scattering centers. This is the Debye mass given by $\mu_D \approx 3T$. The $2 \rightarrow 2$ cross-section then takes the form:

\begin{equation}
\sigma_i (E,T) = \int {\rm d}|\hat{t}| \int {\rm d}x \sum_{j \in \{ q,\overline{q},g \}} f_{j}^{i} (x,\hat{t}) \frac{{\rm d}\hat{\sigma}_j}{{\rm d} \hat{t}} (x\hat{s}, |\hat{t}|)
\end{equation}

Where $\hat{s}$ is the collision energy on the center of mass, $\hat{\sigma}$ is the $2 \rightarrow 2$ scattering cross-section. And $f_j^i(x,\hat{t})$ is the PDF that describes the probability of finding a parton $j$ in a parton $i$ with momentum fraction $x$ at the momentum transfer $\hat{t}$. The PDF shown in the above equation represents the initial-state radiation emitted by the projectile. This implements higher order effects in the perturbative calculation. In this prescription, Bremsstrahlung is automatically taken into account.
\par
The main difficulty in implementing Monte-Carlo simulations for quantum systems is the fact that quantum systems are not Markov chains. There is always the presence of interference effects that must be taken into account. JEWEL does it so using a parametrically formation time attributed to the radiated gluons from the projectile. If the formation time of emission is longer than two consecutive scatterings, then these cases must be analyzed and dealt with specially.
\par
To summarize, the steps taken by JEWEL are the following:

\begin{itemize}
\item[1] First, a hard scattering is generated in PYTHIA, producing the partons to be propagated;
\item[2]JEWEL evolves this parton state with collisions and radiation, taking into account the non-abelian Landau-Pomeranchuk-Migdal effect;
\item[3]Once the partons are close enough to mass shell and have escaped the medium, the final partonic state is handed back to PYTHIA;
\item[4]PYTHIA performs the fragmentation of the partonic final state and hadronization, as well as the resonance decays;
\end{itemize}

\begin{figure}
\begin{minipage}{0.1\textwidth}
$2 {\rm Re}\Bigg[ \Bigg($
\end{minipage}
\begin{minipage}{0.3\textwidth}
\includegraphics[width=1\textwidth]{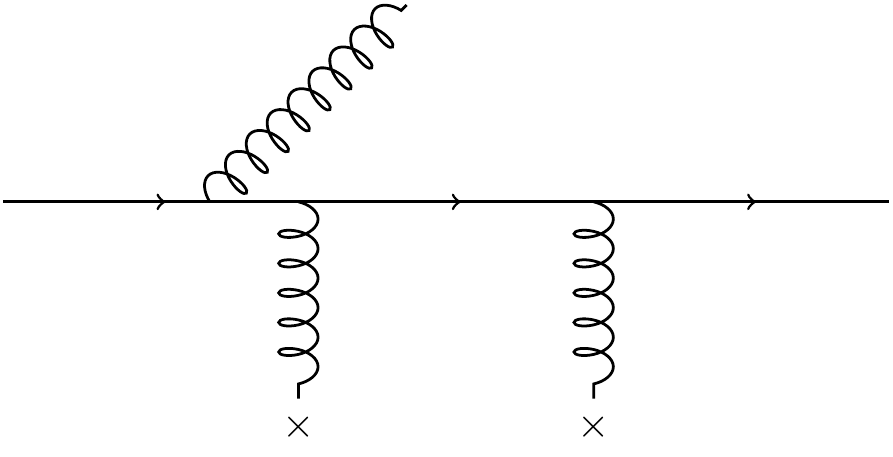}
\label{lpm_a}
\end{minipage}
\begin{minipage}{0.1\textwidth}
$\Bigg) \times \Bigg( $
\end{minipage}
\begin{minipage}{0.3\textwidth}
\includegraphics[width=1\textwidth]{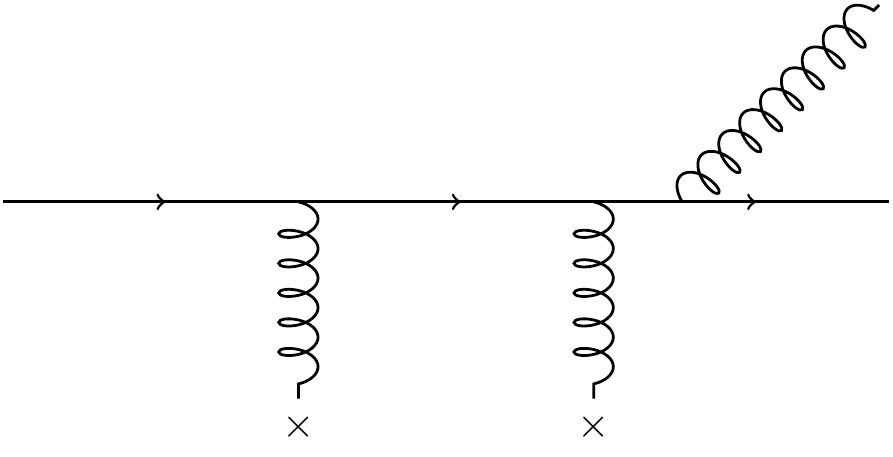}
\label{lpm_b}
\end{minipage}
\begin{minipage}{0.05\textwidth}
$\Bigg)^{*} \Bigg]$
\end{minipage}
\caption{Interference term appearing in radiation stimulated by multiple scattering.}
\label{lpm}
\end{figure}

The LPM effect is an effect that enters multiple scattering when the scale of a radiation does not allow one to treat two successive scatterings as independent. Crossed terms in the squared amplitude, such as the one displayed in Figure \ref{lpm}, create a destructive interference. To handle it, JEWEL proceeds with a formation time prescription that dictates if the successive scatterings are seen as one by the radiated gluon.

\mysection{Hadronization}

Jets, as described before, are a natural consequence of confinement. The formation of jets from partons is called hadronization. Unfortunately, hadronization is a low energy process and falls within the realm of non-perturbative Quantum Chromodynamics. As an implication, a lot of models are used to describe this process and not first principle calculations. One such model is the Lund model. This is the model implements by PYTHIA, which was used to convert the final partonic state provided by JEWEL into a set of hadrons.

\mysubsection{Lund Model}

The Lund string model is based on the idea of tunneling in quantum mechanics. The equation that must be satisfied by the wavefunction of a quark coming out of the Dirac sea is\cite{wong_introduction_1994}:

\begin{equation}
\left[ (p-A)^2 -m^2 \right] \psi = 0
\end{equation}

Where $p$ is the particle 4-momentum, $A$ is the four potential and $m$ is the particle mass.The potential between a quark and an antiquark is linear in the $z$ coordinate. Expanding this equation and separating the variables, we can arrive at:

\begin{equation}
\psi = \exp [ i(p_x x + p_y y -Et) ] f(z) = 0
\end{equation}

Where $f(x)$ satisfies the equation:

\begin{equation}
\left\{ [E - A_0(z)]^2 - p_z^2 - m_T^2 \right\} f(z) = 0
\end{equation}

Where $m_T=\sqrt{p_T^2 + m^2}$. Here we have assumed only a scalar potential $A_0$\cite{wong_introduction_1994}. Dividing this equation by $m_T$ we arrive at:

\begin{equation}
\left[ \frac{p_z^2}{2m_T} + V_{eff}(z) - E_{eff} \right] f(z) = 0
\end{equation}

Which is a Schrödinger equation for an effective energy $E_{eff}=0$ and an effective potential of the form:

\begin{equation}
V_{eff}(z) = \frac{m_T}{2} - \frac{[ E-A_0(z) ]^2}{2 m_T}
\end{equation}

\begin{figure}
\includegraphics[width=0.5\textwidth]{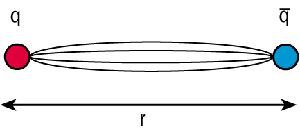}
\caption{Schematics of a quark and an antiquark and the field between them.}
\label{lund_scheme}
\end{figure}

We assume the potential to be linear. To see why this is the case, one can look at Figure \ref{lund_scheme}. The lines of force do not spread into space. This happens because the gluon field carries charge and it is attracted to itself. The gluon field also satisfies Gauss's law. As a result, the flow in any cross-sectional surface between the quarks must be the same. This implies a constant electric field, which in turn, implies a linear potential:

\begin{equation}
A_0(z) = -\kappa z
\end{equation}

Coupling this formalism with the WKB method, we arrive at\cite{wong_introduction_1994}:

\begin{equation}
P = \exp \left\{ - \frac{\pi m_T^2}{\kappa} \right\}
\end{equation}

This equation indicates that particles with a greater transverse mass have a lower probability of being produced and explains the hadrochemistry observed in jets. The mechanism just explained is the Schwinger particle production mechanism. It was created to describe pair creation in strong electric fields in QED. It was adapted to the context of QCD where it could describe the behavior of jets, e.g. their multiplicity.
\par
Once the parton pairs no longer have enough energy to break the strings, they become resonances that decay into hadrons in a Lorentz invariant way. At this stage, phase space considerations are used to determine the momenta of the decay products\cite{sjostrand_pythia_2006}.

\mychapter{Method} \label{method}

\mysection{The simulation}

The simulation built in this work to study the phenomenon of Jet Quenching is performed by coupling JEWEL with an alternative. A description of the models used in this work can be found in Chapter \ref{theory}. First, a simulation was performed with the JEWEL using the default settings, which simulates an idealized version of the initial conditions and the medium\footnote{see Subsection \ref{glauber}}, with the density decaying due to longitudinal expansion only\footnote{see Subsection \ref{bjorken}}. After that, JEWEL code for the medium was modified with an implementation to read the temperature profiles of an arbitrary model. This allowed us to study the effects of different initial conditions and also a realistic hydrodynamic evolution. For this, a choice was made to create a grid with spacing $0.15 \,\rm fm$ on which the temperature profile was read from a foreign model. The time step used is of $0.1 \,\rm fm/c$. This is a time step larger than necessary for hydro simulations, but we are not integrating differential equations, so this does not insert great numerical errors, for more detail refer to the discussion on the last session of this chapter. For intermediate values, a bicubic interpolation was used.
\par 
After this procedure, the generated events on JEWEL were analyzed with code developed with the Rivet\cite{bierlich_robust_2019} package for analysis of the generated Monte-Carlo events. FastJet\cite{cacciari_fastjet_2012} package was also used. The observables generated were the ones in Section \ref{algorithms}. The uncertainty on the histogram bins presented in Sections \ref{jewel_with_ic} and \ref{jewel_with_hydro} are calculated for a binomial distribution according to the statistics generated by the Monte-Carlo events. Since there are specific features for this analysis, such as the recoiling scattering centers\footnote{see Section \ref{background}}, the analysis was built with the aforementioned packages to attend our specific needs.
\par
The simulation in each of the cases consists of a thousand of different medium profiles, and ten thousand JEWEL Monte-Carlo events generated for each profile. The different simulated scenarios are displayed in Table \ref{scenarios}. The analysis was performed in the $0-10\%$ centrality class for $\sqrt{s_{NN}} = 2.76 \, {\rm TeV}$ collision energy. In the case of charged jet observables, scaling factors were extracted from proton-proton collisions, both on the jet energy and also on spectrum normalization\cite{zapp_geometrical_2014}. These factors allow us to compare the analysis of full jets and charged jets. We extract them by matching the spectra of a given observable in pp collisions by scaling it. It is then assumed that the factors are the same in heavy-ion collisions. An example is the jet $p_T$, in this case, since charged jets have fewer particles, its momentum will be smaller, and scaling for comparison is necessary. A factor of $3/2$ was obtained in \cite{zapp_geometrical_2014}.

\begin{table}
\centering
\begin{tabular}{| c | c | c |}
 \hline 
 \multicolumn{3}{| c |}{Scenarios} \\
 \hline
 \hline
            & Initial conditions & Evolution \\ 
 \hline
 Default              & Ideal   & \multirow{3}{10em}{Longitudinal expansion} \\ 
 \cline{1-2}
 \trento              & \trento & \\
 \cline{1-2}
 MC-KLN              & MC-KLN & \\
 \hline
 v-USPhydro + MC-KLN  & MC-KLN  & 2+1 v-USPhydro code \\
 \cline{1-2}
 \hline
\end{tabular}
\caption{Scenarios simulated in this work. See Charter \ref{theory} for explanation on the items of the table.}
\label{scenarios}
\end{table}

To include the new medium profile in JEWEL, a choice was made to build a grid with local values of the temperature. The spacing of the grid is about $0.15 \, {\rm fm}$. The time-step was taken to be $0.1 \, {\rm fm/c}$. The information was then passed to JEWEL upon request through a function $T(x,y,\tau)$. The $\tau$ coordinate is the proper-time and the $x$ and $y$ coordinates are the transverse coordinates, relative to the beam axis. $\tau$ was taken to be equal $\sqrt{t^2-z^2}$ according to a boost invariance assumption. If one chooses not to make this assumption, a 3+1 code simulation for the hydrodynamics is necessary. The restriction to mid-rapidity of all analysis was then used. For values outside the gridpoints, a bicubic interpolation was used\cite{vetterling_numerical_1992}, and on the proper-time coordinate, a linear interpolation was used.
\par
Since only a temperature profile was provided to JEWEL, a choice of an EOS had to be used to provide a density of scattering centers profile. This was used through an ideal equation of state. The density of scattering centers was taken as $n \propto T^3$. This is the dependence as one would have for the formalism described in Section \ref{qgp}. Also, no local fluid velocity was assumed. Any momentum that the scattering centers might have comes from a kinetical description.

\mysection{Jet Algorithms}
\label{algorithms}

\begin{figure}
\includegraphics[width=0.8\textwidth]{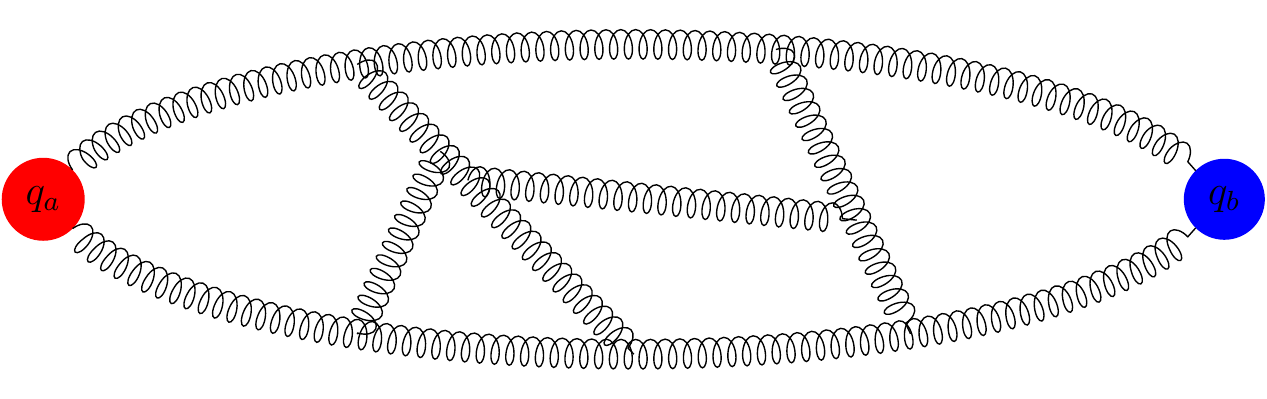}
\caption{Schematics of force lines created by gluon trajectories between quarks.}
\label{force_lines}
\end{figure}

Jets are a consequence of the confinement property of strong interactions\cite{halzen_quarks_1984,peskin_introduction_1995,salam_towards_2010}. It happens because the energy of partons coming out of a hard scattering grows linearly with their distance. This makes it energetically favorable for the system to convert this color dipole into a multi particle state of color neutral hadrons. One common picture of this is that the field between a parton dipole is seen as a tube connecting them. It is a consequence of the self-interaction of the gluon field that the lines of force do not spread out into space, but are contained in the space between the partons, this situation can be seen in Figure \ref{force_lines}. Once the distance between the partons starts to grow, the energy stored in this flux-tube starts to grow linearly. Eventually, it becomes favorable for the system to exchange a portion of this tube for a pair of partons pulled out of the vacuum, turning it into two pairs of partons with two chromoelectric tubes. The process continues repeatedly until the energy in each pair is no longer enough to break the strings. At this point, these less energetic systems decay into the known hadron states.

Armed with the qualitative idea of what a jet is, a definition must be made in a more quantitative ground. This definition must account for certain conditions that are defined in the Snowmass accord\cite{salam_towards_2010}:

\begin{itemize}

\item Simple to implement in an experimental analysis;
\item Simple to implement in the theoretical calculation;
\item Defined at any order in perturbation theory;
\item Yields finite cross sections at any order of perturbation theory;
\item Yields a cross-section that is relatively insensitive to hadronization;

\end{itemize}

The difficulties that the previous accord tried to address are related to the fact that jets are ill-defined objects. There is a major uncertainty because jets are a consequence of non-perturbative physics, which is poorly understood. The way one defines a jet, then, is by defining an algorithm that clusters particles into jets. With this algorithm, experimental analysis can be performed. On the theory side, one usually implements a model for fragmentation and hadronization and later applies the jet algorithm to the final state to make comparisons with data. There is also the alternative to invoke parton-hadron duality and claim the jet 4-momenta are directly comparable to partons momenta\cite{dissertori_quantum_2003}.
\par
Building an algorithm might have some difficulties due to infinities that appear in calculations at the theoretical level. Some of these infinities might be washed away from the calculations by grouping different final states together. The problem, then, is that the algorithm must treat these final states on the same ground as well.
\par
\begin{figure}
\includegraphics[width=1.0\textwidth]{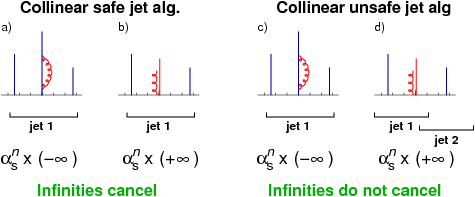}
\caption[Collinear safe ilustration]{A description of collinear safe and an collinear unsafe algorithm. Here we have two situations that must be treated on equal grounds due to theoretical reasons and might not be so due to the choice of jet algorithm. Figure from \cite{salam_towards_2010}}
\label{collinear_safe}
\end{figure}

One of these situations arises when a parton splits collinearly. This situation is illustrated in Figure \ref{collinear_safe}. The loop diagram for the case that the partons merge again cancels its divergence with the case in which the partons split and do not merge. This means that, at the theoretical level, the two cases must be treated at the same level. In Figure, we see the difference between theoretically safe and an unsafe algorithm. An unsafe algorithm might see a hard seed on the high transverse momentum particle on the left and cluster from there. This would result in two jets. A safe algorithm would not be subject to this problem.

\begin{figure}
\includegraphics[width=1.0\textwidth]{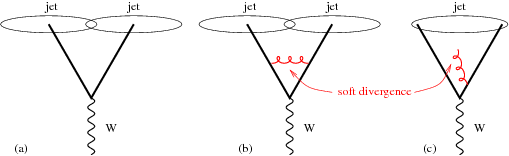}
\caption[Collinear safe ilustration]{A description of infrared safe and an infrared unsafe algorithm. Here we have two situations that must be treated on equal grounds due to theoretical reasons and might not be so due to the choice of jet algorithm. Figure from \cite{salam_towards_2010}}
\label{infrared_safe}
\end{figure}

Another type of problem that might occur is due to soft radiation. This situation can be seen in Figure \ref{infrared_safe}. As before, we have two final states that must be treated on the same ground, from the theoretical perspective. But the algorithm might not do so. It can happen because the extra particle seen in red in the Figure might act as a new seed and cluster two jets into one. Algorithms that succeed in avoiding this merging are called infrared safe.

One of the algorithms that satisfy to a good extent these criteria is the anti-kt algorithm\citep{cacciari_fastjet_2012}. The anti-kt algorithm belongs to a broader class called the recombination algorithms. The members of this class are all based on the idea of a distance defined between every pair of particles being analyzed. If the distance is smaller than the distance of another particle to the beam, the particles are combined. This process is repeated until no single pair satisfies the condition and one ends up with a set of jets of the event. At this stage, further cuts might be done in the rapidity or the $p_t$ of the jet. The distance used in the anti-kt algorithm is defined by the following equation:

\begin{equation} \label{jet_alg}
d_{ij} = \min (p_{t i}^{2p},p_{t j}^{2p}) \frac{\Delta R_{ij}}{R}
\end{equation}

Where $R$ is a parameter that is defined at each analysis. For $p=-1$\footnote{The value $p=1$ corresponds to the $k_t$ algorithm and $p=0$ to the Cambridge-Achen. See \cite{salam_towards_2010} for further details}. Where:

\begin{equation} \label{delta_r}
\Delta R_{ij} = \sqrt{ (\eta_i - \eta_j)^2 + (\phi_i - \phi_j)^2 }
\end{equation}

A distance to the beam axis is also defined:

\begin{equation}
d_i = p_{ti}^{-2}
\end{equation}

\begin{figure}[h]
\includegraphics[width=0.8\textwidth]{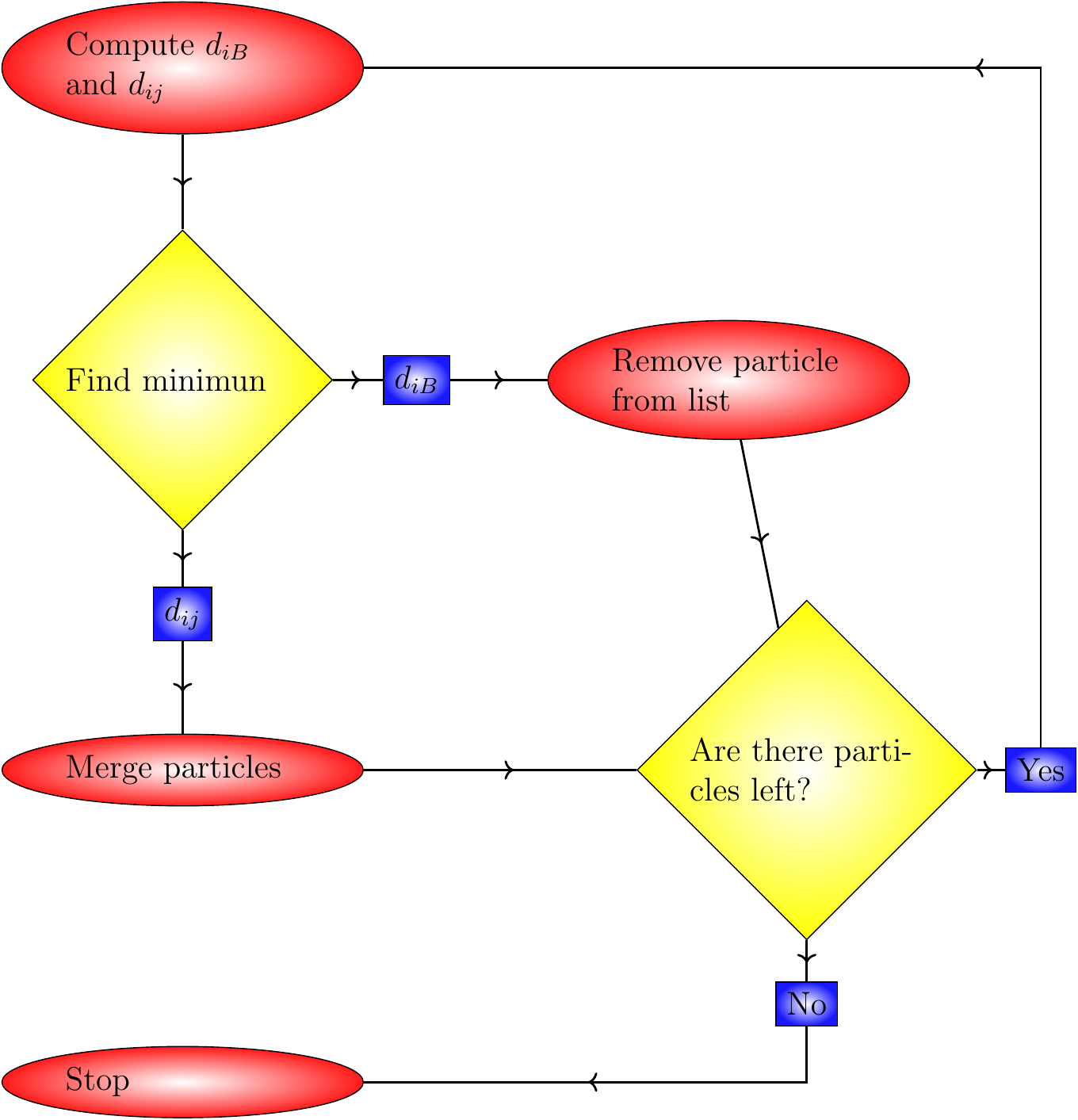}
\caption{Flow diagram for the anti-kt algorithm.}
\label{anti_kt_flow}
\end{figure}

From all the $d_{ij}$ and $d_i$ the minimum value is chosen. If it is $d_i$ then this particle is removed from the list and called a jet. Otherwise, it is a distance $d_{ij}$, the $i$ and $j$ particles are combined. The process goes on iteratively until a final list of jet candidates is left. The recombination of the particles is performed using the $E$ scheme, or 4-momentum scheme, which means particles are combined by adding their 4-momenta. This process is ilustrated in the Figure \ref{anti_kt_flow}.

\mysubsection{Background subtraction}
\label{background}

An extra difficulty arises when trying to perform jet analysis in heavy-ion collisions due to the background contamination, called the Underlying Event (UE). The UE corresponds to the soft particles that come from the radiation of the freeze-out surface and are mainly explained through hydrodynamics. When one is trying to cluster jets, experimentally, one must subtract this background in other to make comparisons with theory.
\par
In JEWEL, the recoiling scattering centers can be kept and included in the final state of the events. Whenever there is an elastic collision during the parton propagation, the distinction between the 4 momentum that is part of the jet and the 4 momentum that is part of the medium is spoiled. To address this problem, the scattering centers kept by JEWEL are used to subtract this 4 momenta from the final state.
\par
Following \cite{zapp_geometrical_2014}, the background subtraction is performed by the use of ghost particles. These particles are inserted into the final state with very low $p_T$ and the same direction as the four-momentum assigned to the scattering center. Particles are then classified as background if they satisfy the following condition:

\begin{equation}
\Delta R_{ij} < 1 \cdot 10^{-5}
\end{equation}

Where $\Delta R_{ij}$ is the distance to the closest ghost particle (see Equation \eqref{delta_r}). The list compiled by evaluating this condition is then subtracted according to each observable. In the case of jet girth, for instance, the particles deemed to be background are removed from the jet constituents list and thus do not participate of the observable calculation. In the case of jet dispersion, we used:

\begin{equation}
p_T^D = \frac{\sqrt{ \sum_{\rm particles} {p_t}^2 - \sum_{\rm background} {p_t}^2 }}{p_{t,corrected}^{jet}}
\end{equation}

For the jet mass, a correction of the jet four-momentum comes automatically when removing the background from the list of particles.

\mysection{Jet Observables}

Although the main idea of a jet is to reconstruct the basic partonic kinematics of a hard scattering, there are a lot of things that can be investigated with a jet. This is since the jet has more information than a single particle. One can study its shape and structure to find radiation patterns and non-perturbative dynamics. In the present work, the main idea is to look for radiation patterns. Here we define the observables that can characterize the jets used in this work.

\mysubsection{Girth}

The generalized angularities for the jet constituents are defined as:

\begin{equation} \label{gen_ang}
\lambda_{\beta}^{\kappa} = \sum_i \left( \frac{p_{T,i}}{p_{T,jet}} \right)^\kappa \left( \frac{\Delta R_{jet,i}}{R} \right)^\beta
\end{equation}

Where $\kappa$, $\beta$ and $R$ are parameters chosen at each analysis. $\Delta R_{jet,i}$ is the angular distance between jet and constituent $i$. Girth is related to the angular opening of the jet. The idea is to pick the first order moment in angular distribution with the constituents $p_T$ as weights. This means picking $\beta=\kappa=1$. Girth is defined by:

\begin{equation}
g = \sum_{\rm constituents} \frac{p_t}{p_t^{jet}} \Delta R_{i,jet}
\end{equation}

Where $\Delta R_{i,jet}$ is calculated according to \eqref{delta_r} and the sum is taken on the constituents of the jet. Its possible values lie in the $\left[ 0,1 \right]$ interval. Girth is also known in some collaborations as jet width. Unlike the jet mass, it depends linearly on the transverse momentum of the constituent particles. Therefore, it is more sensitive to softer fragmentation.

\mysubsection{Dispersion}

This observable measures the \emph{hardness} of the fragmentation of the jet. This means that it will have a large value if the jet $p_t$ is distributed in fewer particles, and smaller values if the $p_t$ is distributed in a larger number of particles. Its definition is given by the equation:

\begin{equation}
p_T^D = \frac{\sqrt{ \sum_{\rm constituents} {p_{t,i}}^2 }}{p_t^{jet}}
\end{equation}

Where the sum is taken on the constituents of the jet. It corresponds to $\kappa=2$ and $\beta=0$ in Equation \eqref{gen_ang}. It can be seen from its definition that the possible values lie in the $\left[0 , 1 \right]$ interval. Values close to one will correspond to the extreme case of the jet transverse momentum being distributed in one or two constituents and one of them carries most of its momentum. The opposite limit corresponds to the jet transverse momentum being distributed almost equally on a high number of constituents.

\mysubsection{Jet Mass}

The jet mass is a observable that is constructed from the jet four-momentum in the usual Lorentz formalism. The definition is:

\begin{equation}
M_J = \sqrt{ \left( \sum_{\rm particles} {\rm E} \right)^2 - \left( \sum_{\rm particles} {\rm \overrightarrow{p}} \right)^2 }
\end{equation}

It can be seen, in the RHS of the above equation that this observable is sensitive to the collimation of the jet. This can be observed by noting that, inside the square root, there is a scalar sum and a vector sum, where vector sum depends on the angular separation of the constituents. The broader the jet, the smaller the jet mass. And the impact of each constituent particle in this observable depends on the squared transverse momentum. This increases the impact of the harder constituents of the jet.





\mysubsection{Azimuthal anisotropy $v_2$}

The azimuthal anisotropy for a given observable comes from the idea of expanding a given spectrum in the following form:

\begin{equation}
\frac{{\rm dN}}{{\rm d}^2p_T {\rm d}\eta {\rm d}\phi } = \frac{{\rm dN}}{{\rm d}^2p_T {\rm d}\eta } \left[ 1 + \sum_n 2 v_n \cos \left( n ( \phi - \Psi_{\rm RP} ) \right) \right]
\end{equation}

Where $\Psi_{\rm RP}$ is the reaction plane angle. The coefficients $v_n$ are adjusted and represent the lack of cylindrical symmetry of the spectra. $v_2$ is called the azimuthal anisotropy coefficient and $v_3$ is the triangular flow. They usually reflect the geometry of the initial conditions set by the early dynamics, before the hydro phase. The harder problem in the measurement of these quantities is the determination of the reaction plane, which is random. Usually, it is calculated by adjusting the above formula with the distribution in $\phi$ of the observed particles in an event. The $v_2$ coefficient is related to the ellipticity of the initial conditions for the hydro expansion. As can be seen in Figure \ref{jet_v2}, it comes from the shape of the overlap zone of the colliding nuclei.

\begin{figure}
\includegraphics[width=.8\textwidth]{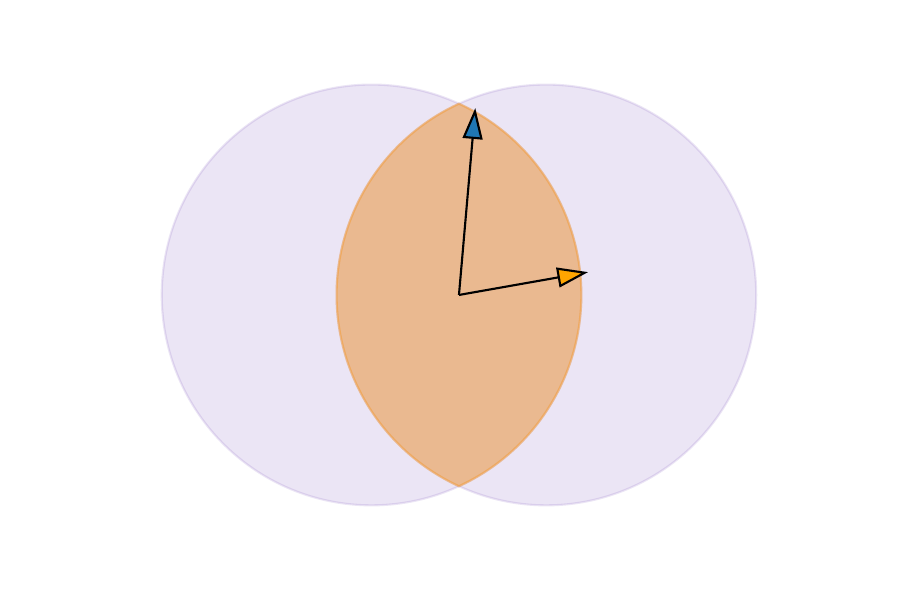}
\caption{Schematics of path length dependence of jet quenching. The blue arrow represents a longer trajectory for a parton, and therefore has a higher chance of losing more energy then the orange arrow.}
\label{jet_v2}
\end{figure}

\mysection{Grid validation}

\begin{figure}
\includegraphics[width=0.7\textwidth]{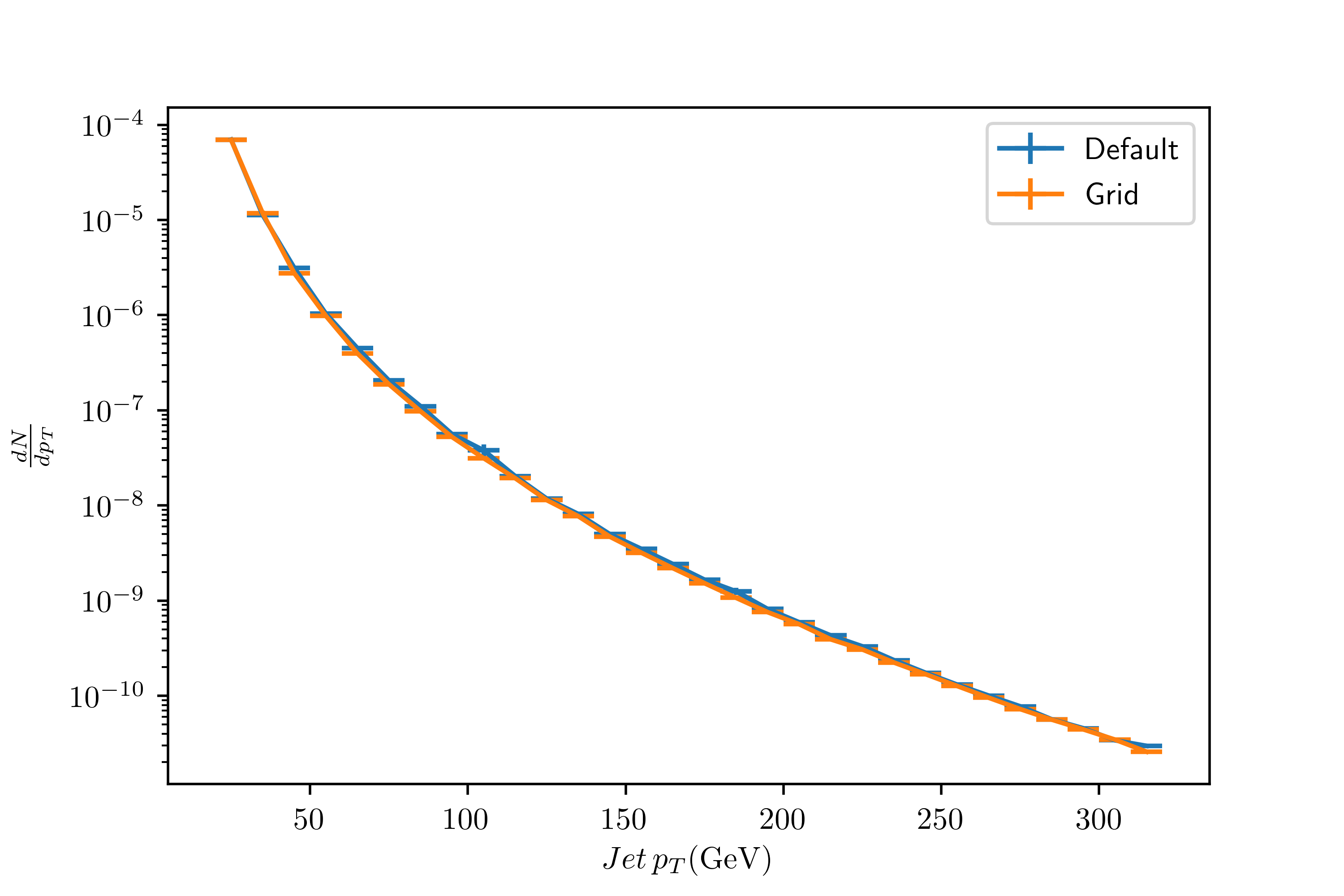}
\caption[Jet $p_T$ for Grid validation.]{Jet $p_T$ for Grid validation.}
\label{grid_jetpt_validation}
\end{figure}

\begin{figure}
\includegraphics[width=0.7\textwidth]{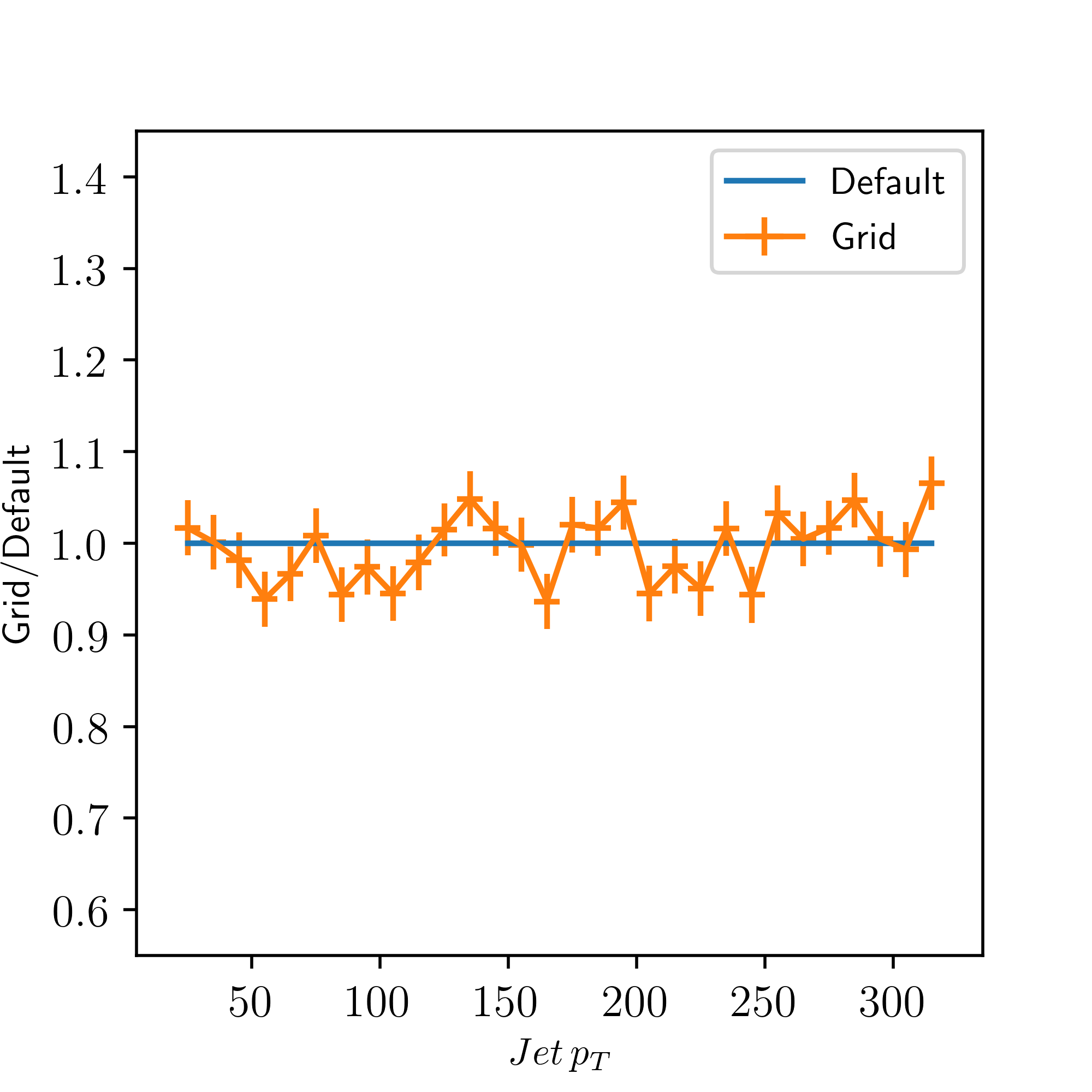}
\caption[Grid/Default for jet $p_T$]{Grid/Default for jet $p_T$}
\label{grid_default}
\end{figure}

As a validation for the grid method implemented in this work, we show here a comparison between observables calculated with JEWEL in its default configuration and with the grid method implemented here. In Figure \ref{grid_jetpt_validation} we can see the jet $p_T$ spectrum for both cases. And in Figure \ref{grid_default} we can see a ratio plot of both spectra. Within the statistical uncertainty calculated for the Monte Carlo, we can see that they both agree. This result shows that errors inserted due to interpolation of the grid do not affect this observable.

\begin{figure}
\includegraphics[width=0.7\textwidth]{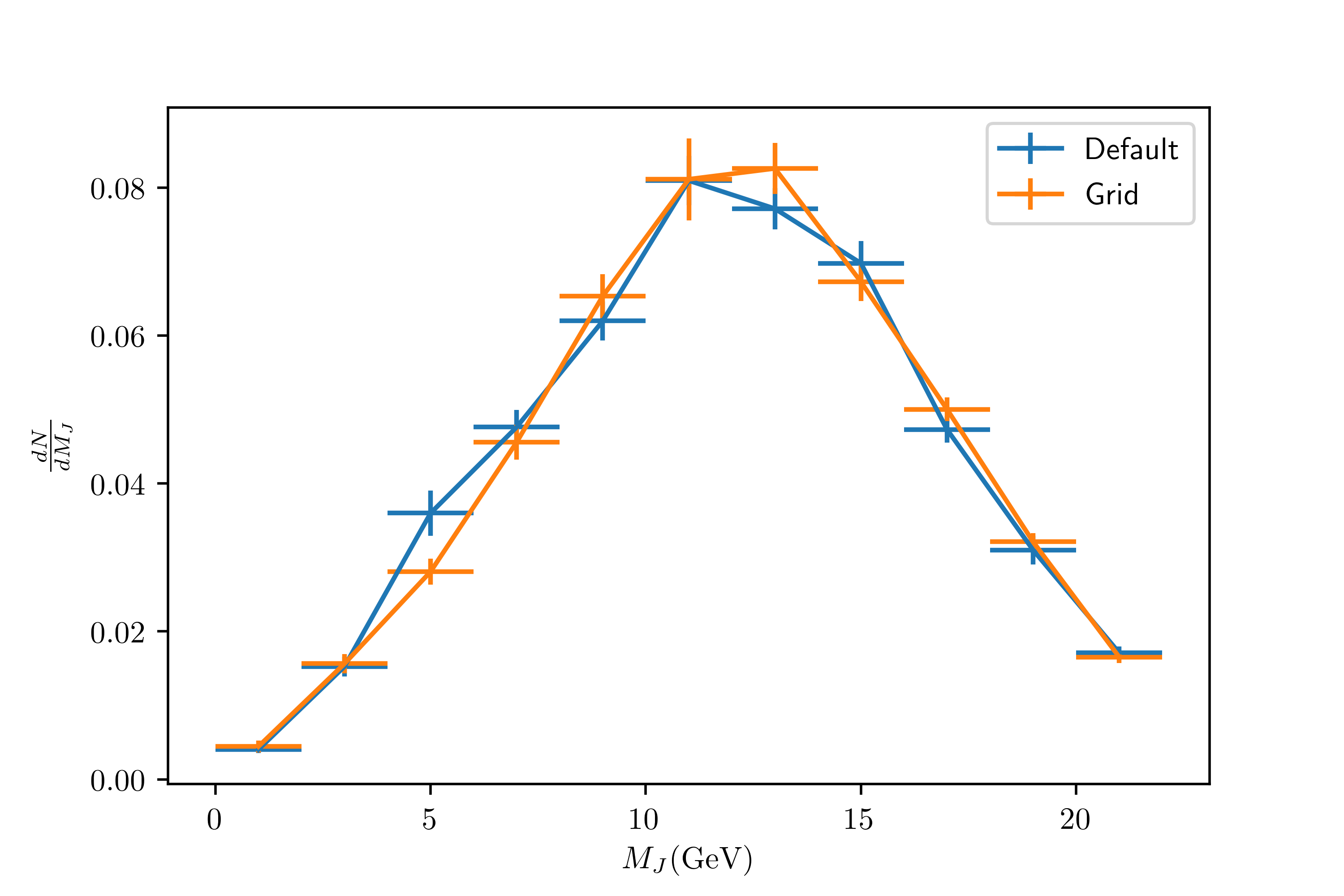}
\caption[Jet Mass for Grid validation.]{Jet Mass for Grid validation.}
\label{grid_jetmass_validation}
\end{figure}

\begin{figure}
\includegraphics[width=0.7\textwidth]{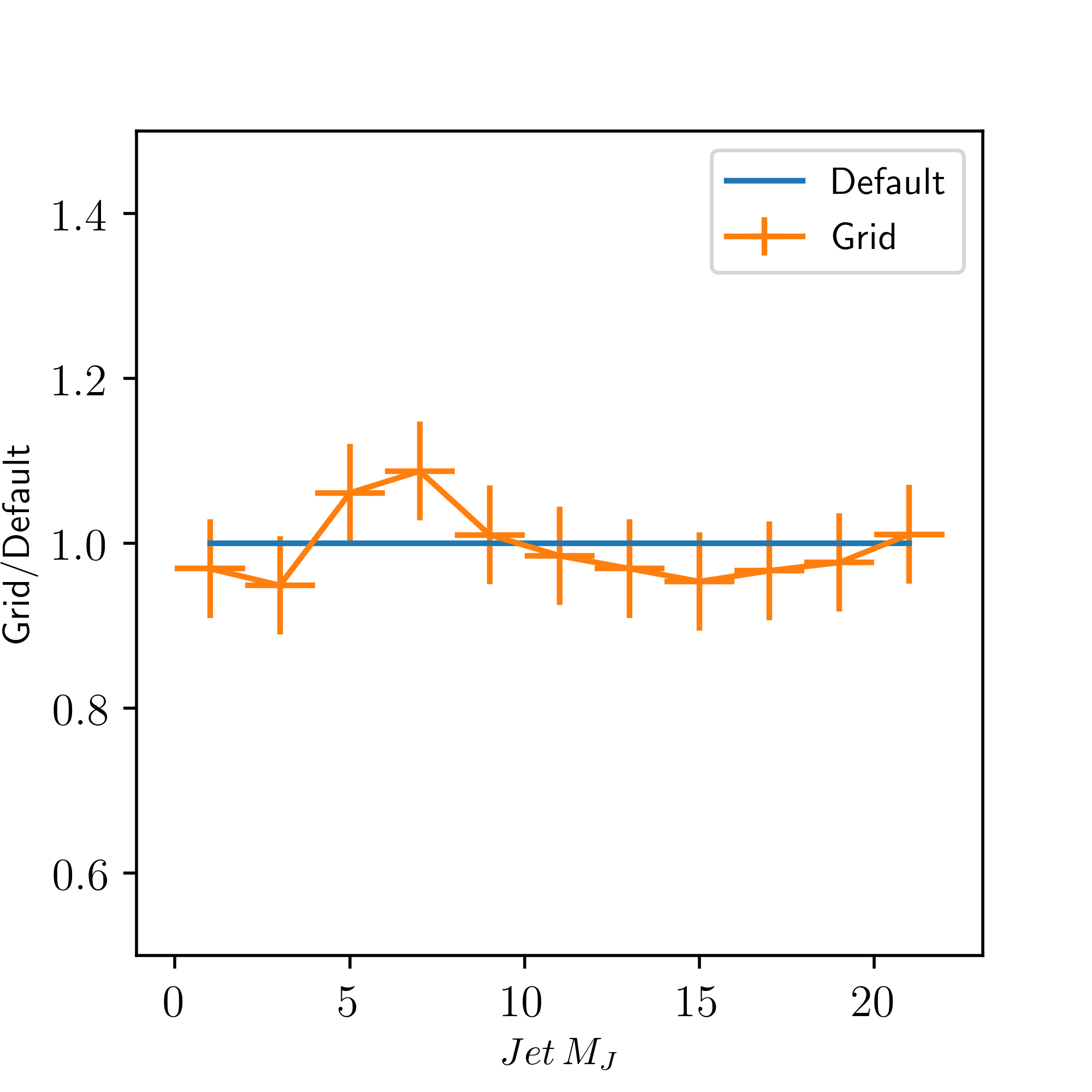}
\caption[Grid/Default for jet mass]{Grid/Default for jet mass}
\label{grid_default_jetmass}
\end{figure}

For the case jet mass we can see a plot of the spectra and a ratio plot on Figures \ref{grid_jetmass_validation} and \ref{grid_default_jetmass}, respectively. The mass has an error of $25\%$ in only one bin, but the rest of the bins is well controlled within the uncertainties. This shows that both methods agree. The peak and width of the distribution are well reproduced.

\begin{figure}
\includegraphics[width=0.7\textwidth]{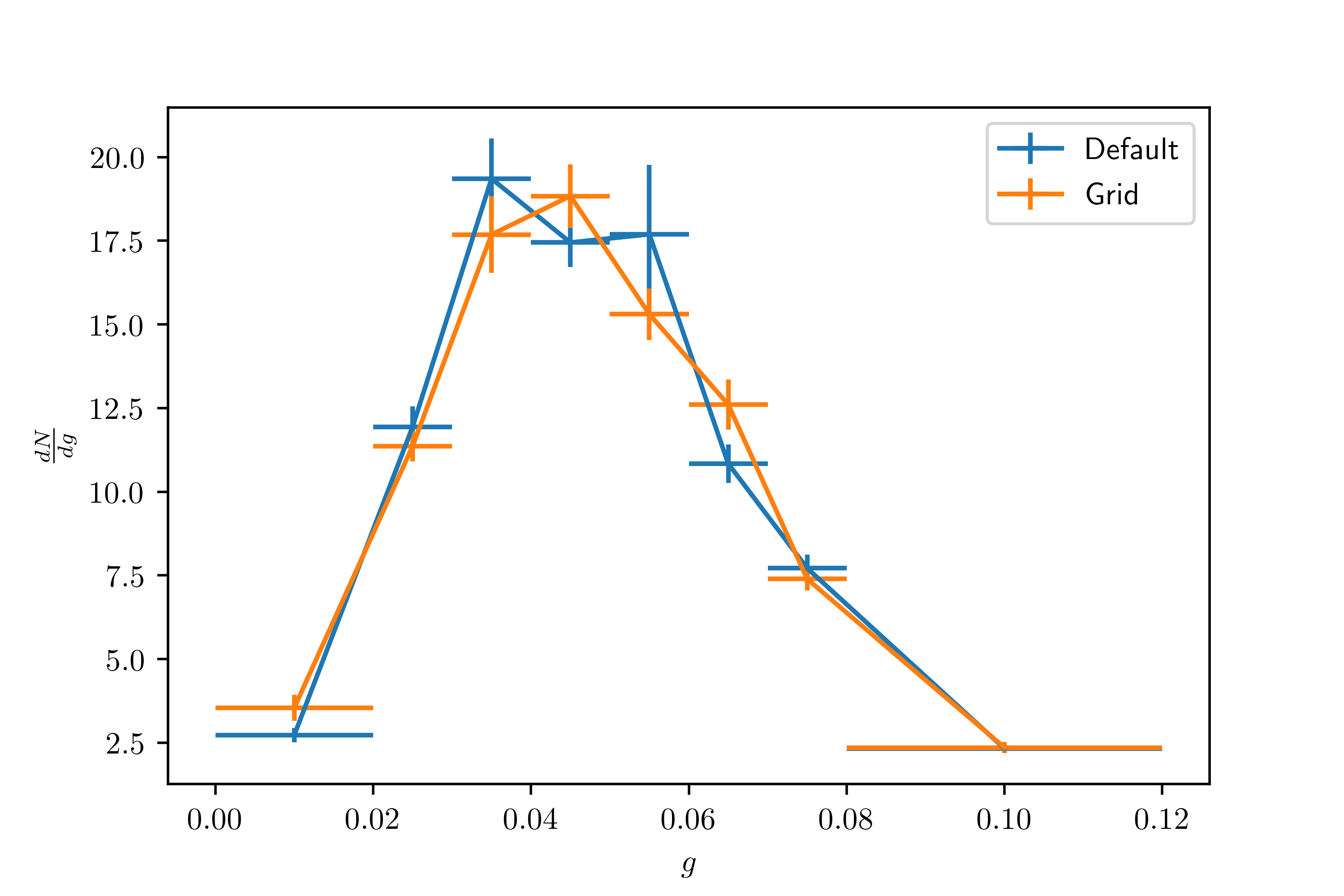}
\caption[Jet Girth for Grid validation.]{Jet Girth for Grid validation.}
\label{grid_girth_validation}
\end{figure}

\begin{figure}
\includegraphics[width=0.7\textwidth]{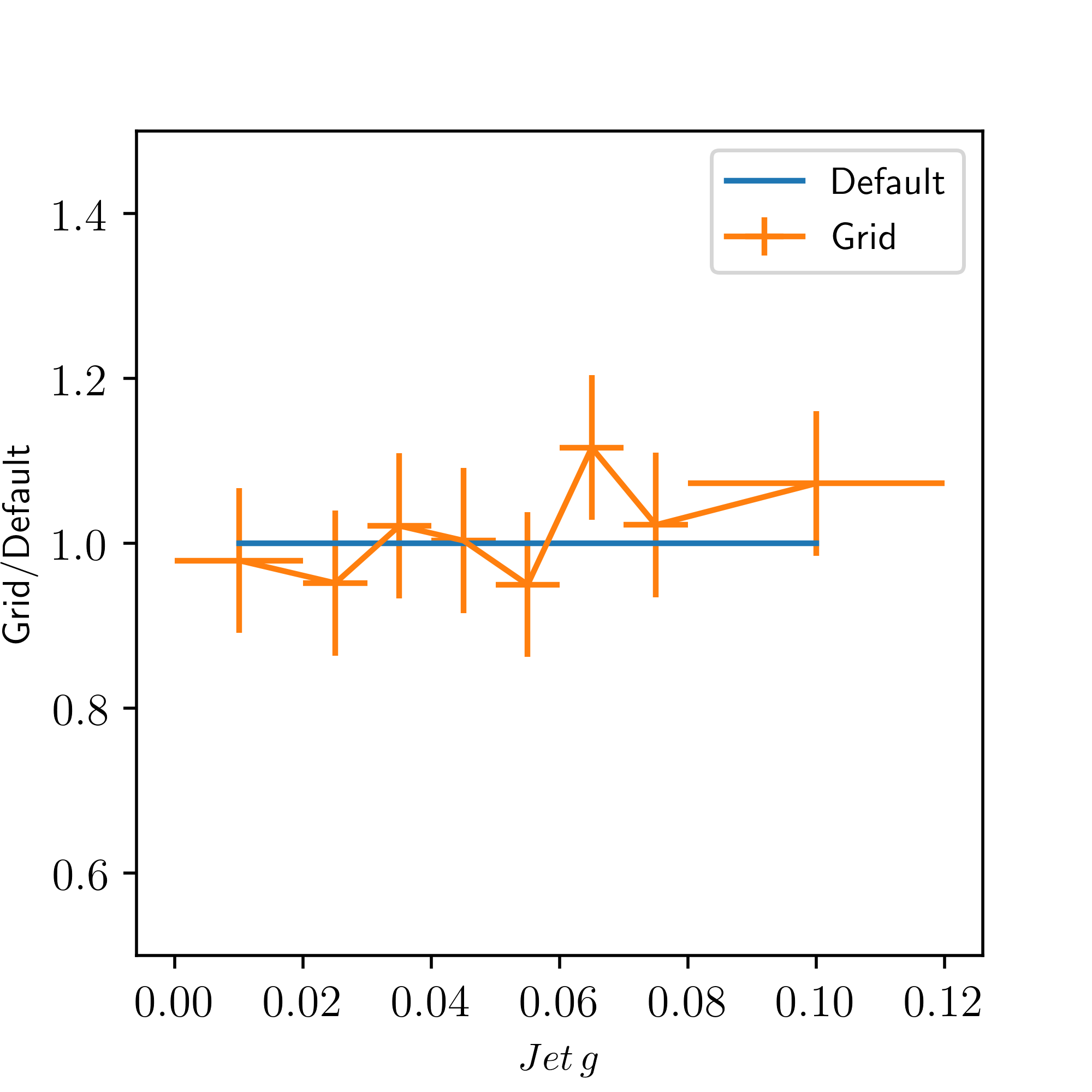}
\caption[Grid/Default for jet girth]{Grid/Default for jet girth}
\label{grid_default_girth}
\end{figure}

On Figures \ref{grid_girth_validation} and \ref{grid_default_girth} we can see the plots for the jet girth for both cases and also a ratio plot. The qualitative behavior is reproduced. Within the uncertainties, the data of both methods agree. The peak and width of the distribution are well reproduced.

\begin{figure}
\includegraphics[width=0.7\textwidth]{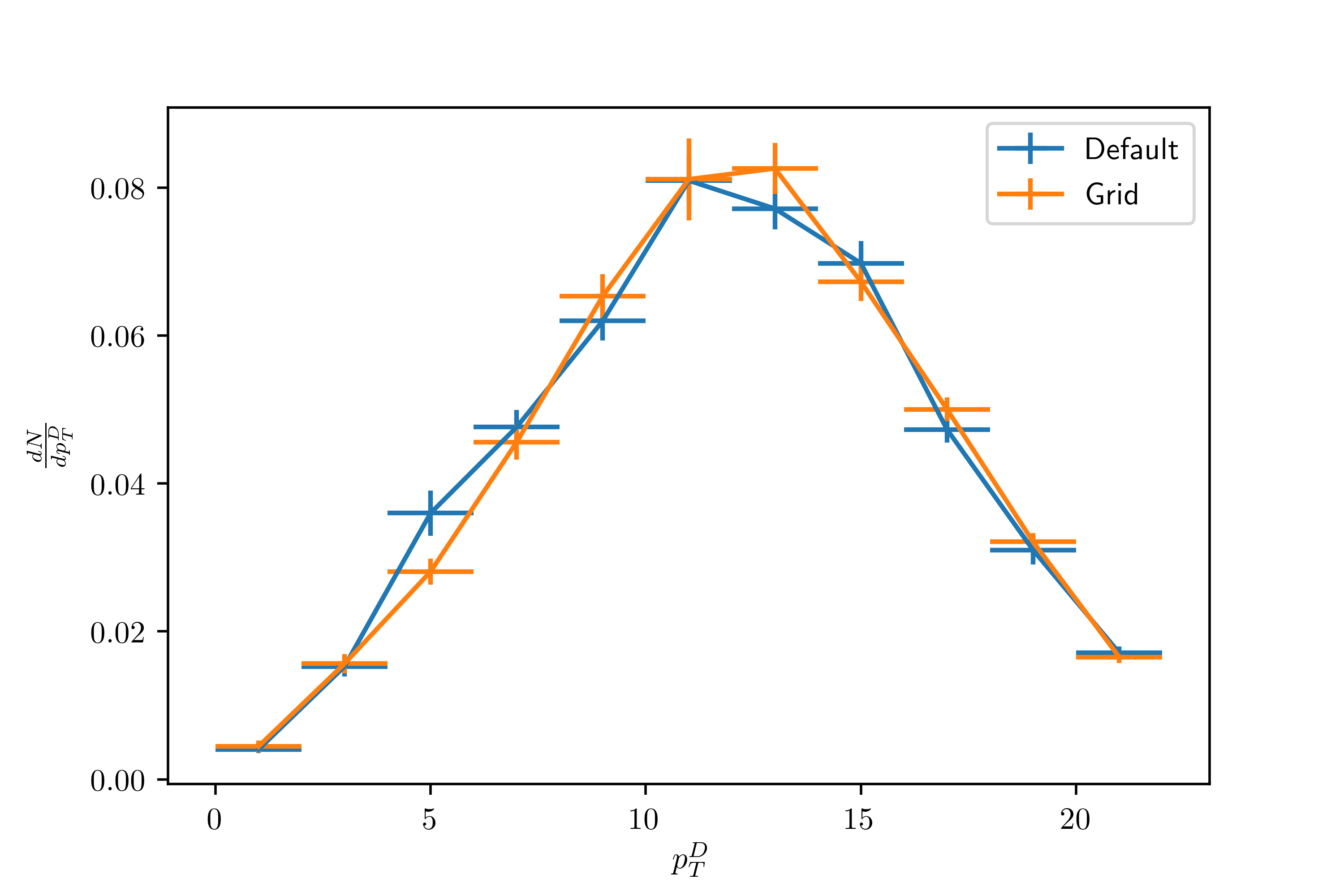}
\caption[Jet Dispersion for Grid validation.]{Jet Dispersion for Grid validation.}
\label{grid_dispersion_validation}
\end{figure}

On Figures \ref{grid_dispersion_validation} and \ref{grid_default_girth} we can see the plots for the jet $p_T^D$ for both cases and also a ratio plot. The qualitative behavior is reproduced. Within the uncertainties, the data of both methods agree. The peak and width of the distribution are well reproduced.

\begin{figure}
\includegraphics[width=0.7\textwidth]{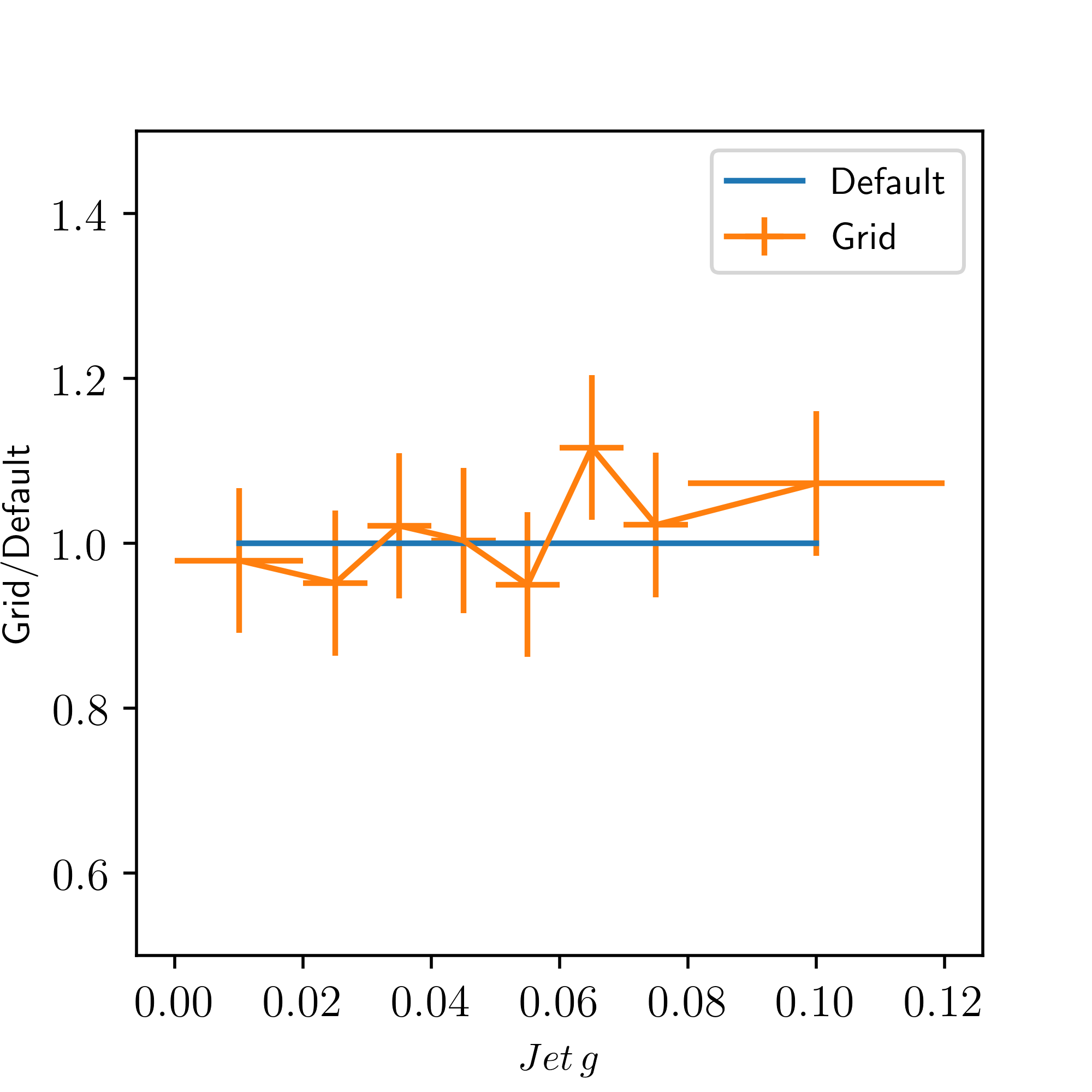}
\caption[Grid/Default for jet dispersion]{Grid/Default for jet dispersion}
\label{grid_default_dispersion}
\end{figure}

\mychapter{Results} \label{results}

\mysection{Experimental Results}

\begin{figure}
\includegraphics[width=0.6\textwidth]{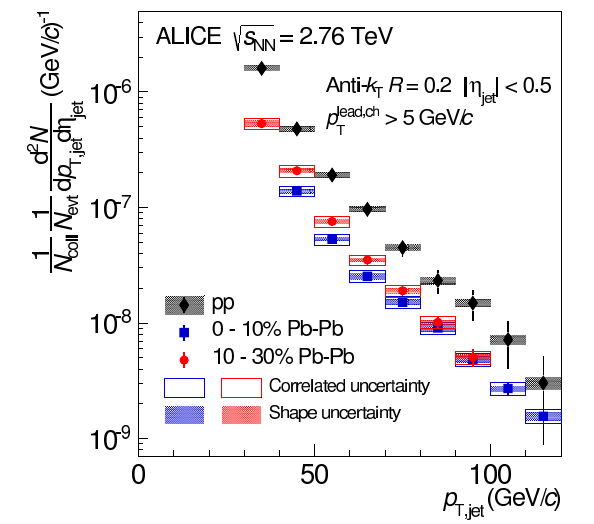}
\caption[Experimental Jet $p_T$]{The spectra of $R = 0.2$ jets with a leading track requirement of $5 \, {\rm GeV/c}$ in $0-10\%$ and $10-30\%$ most central Pb–Pb collisions scaled by $1/N_{coll}$ and in inelastic pp collisions at $\sqrt{s_{NN}} = 2.76 \, {\rm TeV}$. Plot from \cite{alice_collaboration_measurement_2015}}
\label{exp_jet_pt}
\end{figure}

\begin{figure}
\includegraphics[width=0.6\textwidth]{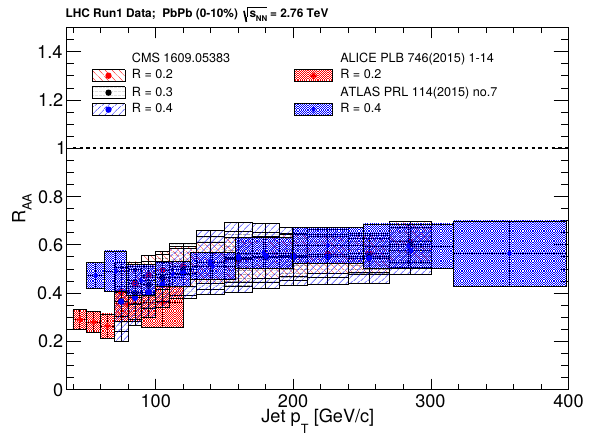}
\caption[Experimental Jet $p_T$ $R_{AA}$]{Jet $p_T$ $R_{AA}$ measured in different collaborations. Figure from \cite{connors_review_2017}.}
\label{exp_jet_pt_raa}
\end{figure}

Before showing the results of the simulations, it is interesting to discuss the experimental results for the studied observables. They indicate that jets suffer modifications due to interactions with the dense and hot medium created in relativistic heavy-ion collisions. The first of these, and most inclusive one is the jet $p_T$. We can observe in Figure \ref{exp_jet_pt} that the PbPb spectrum is suppressed when compared to the pp spectrum. The same trend is observed when central and semi-peripheral collisions are compared: the former is more suppressed than the later. This indicates that there is suppression of jets for PbPb collisions, and that this suppression is also related to centrality. The fact that it varies with the centrality is also an evidence that this suppression has its origin in the interaction with the medium. Higher centrality implies higher energy densities, which in turn implies higher chance of forming the Quark-Gluon Plasma,. In the Figure \ref{exp_jet_pt_raa} we see that the PbPb spectrum can be $20\%$ that of pp for lower transverse momentum, and saturates at no more than $60\%$ for higher values of $p_T$.

\begin{figure}
\includegraphics[width=1.0\textwidth]{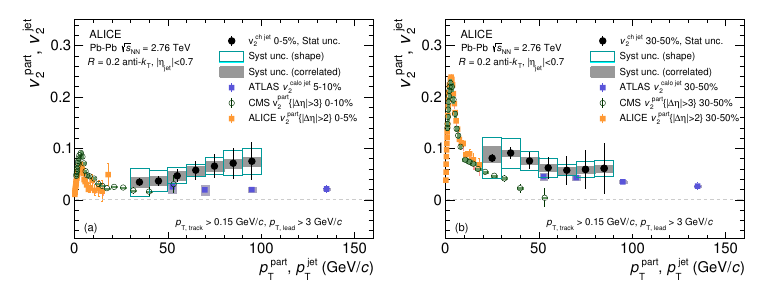}
\caption[Experimental Jet $v_2$]{Jet $v_2$ measured in different collaborations. Figure from \cite{connors_review_2017}.}
\label{exp_jet_v2}
\end{figure}

There is also evidences that this suppression is path length dependent. This can be seen in Figure \ref{exp_jet_v2} where several measurements of $v_2$ are presented. The data in orange and in white circles are measurements of the $v_2$. They come from ALICE and CMS collaborations respectively. The fact that it grows linearly for lower $p_t$ is predicted by modeling collective behavior. For $p_T \gtrsim 5\,{\rm GeV}$, the particles are not usually thermalized. The description of the particles as Jet Quenching then comes into play. We see in the plot that the $v_2$ continues to be non-zero well above $5\,{\rm GeV}$. This indicates that the energy loss of this partons must be path length dependent. The fact that it depends also on centrality is evidence that this comes from the interaction of high energy partons with the medium. The data in black circles comes from ALICE collaboration, and in blue squares come from ATLAS collaboration. This data is different from the previous cases since it uses reconstructed jets. ALICE reconstructs jets with the TPC, so only charged particles are included in the analysis. ATLAS uses the hadronic calorimeters, which means it measures full jets. Measuring only charged particles implies, for the same jet, measuring less particles, which imply less $p_T$ for the same jets. This explains the difference in scale of the reconstructed jets observed in Figure \ref{exp_jet_v2}. For semi-central and central collisions, the collaborations do not agree upon a re-scaling of jet momenta. ALICE measures a higher value for $v_2$.

\begin{figure}
\includegraphics[width=1.0\textwidth]{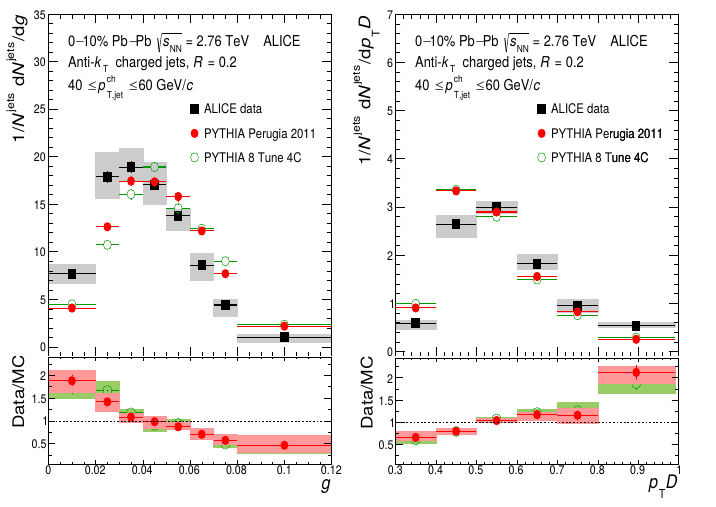}
\caption[Experimental Girth and $p_t^{D}$]{Girth and $p_t^D$ measured by ALICE collaboration. Figure from \cite{alice_collaboration_medium_2018}.}
\label{exp_girth_ptd}
\end{figure}

\begin{figure}
\subfloat[$p_T^D$]{
\includegraphics[width=0.5\textwidth]{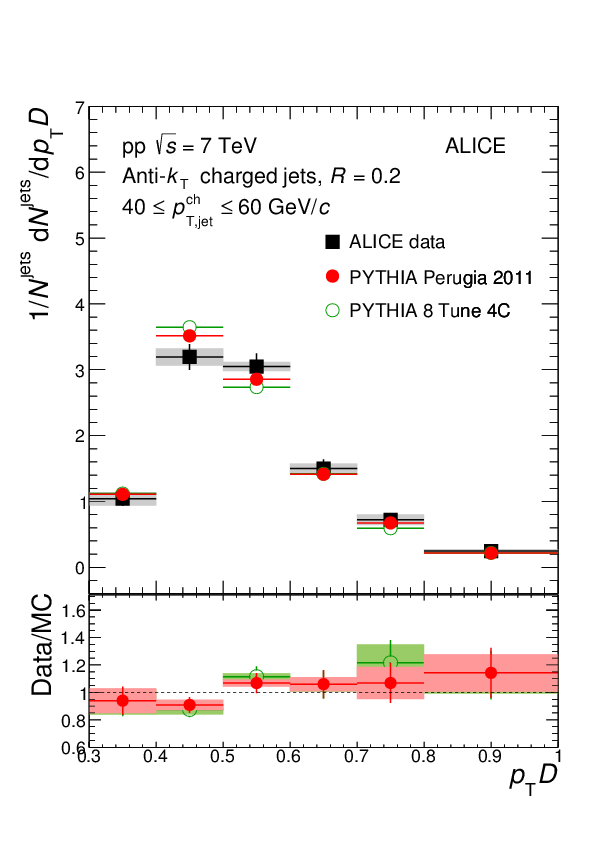}
\label{ex_ptd_pp}
}
\subfloat[$g$]{
\includegraphics[width=0.5\textwidth]{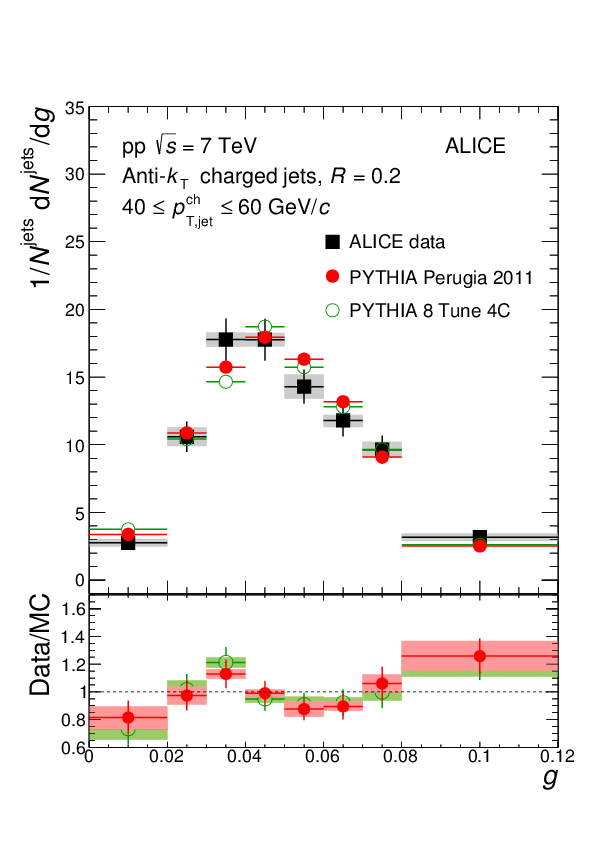}
\label{ex_g_pp}
}
\caption[$p_T^d$ and $g$ for pp collisions.]{Measuraments of both $p_T^D$ and $g$ compared with simulations for pp collisions. Figure from \cite{alice_collaboration_medium_2018}.}
\label{exp_pp}
\end{figure}

In Figure \ref{exp_girth_ptd} we see measurements of the girth and $p_t^{D}$ for charged jets with small radius $(R=0.2)$ jets. The simulations for pp describe well the data for these observables\cite{alice_collaboration_medium_2018}, this can be seen on Figure \ref{exp_pp}. There is a modification if compared with the simulation also displayed in the plot. The girth indicates more collimated jets. The $p_t^D$ indicates harder fragmentation if compared to the pp case.

\begin{figure}
\includegraphics[width=1.0\textwidth]{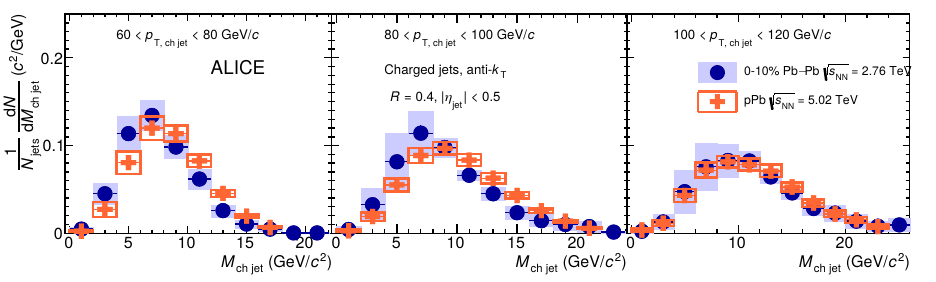}
\caption[Experimental Jet Mass]{Jet mass measured by ALICE collaboration. Figure from \cite{alice_collaboration_first_2018}.}
\label{exp_jet_mass}
\end{figure}

Regarding jet mass, the first measurements can be seen in the Figure \ref{exp_jet_mass}. In the Figure, the mass measured in PbPb collisions is compared to pPb collisions. pp data for jet mass is well described by simulations. In \cite{alice_collaboration_first_2018} the comparison of pPb with simulations for pp show that there are no cold matter nuclear effects on this observable. So the comparison of PbPb with pPb would show only the effects of the hot QGP on the partons. For jets on the range $60-100 \, {\rm GeV}$ range, the jets in PbPb tend to have slightly lower mass than those of pPb. This indicates a broadening of the jet.

\mysection{JEWEL results}

\begin{figure}
\includegraphics[width=0.8\textwidth]{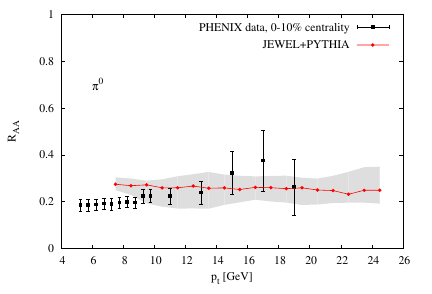}
\caption[Hadron $R_{AA}$]{Hadron $R_{AA}$ as measured by the PHENIX collaboration compared to JEWEL predictions. Figure from \cite{zapp_perturbative_2013}.}
\label{hadron_supression}
\end{figure}

\begin{figure}
\includegraphics[width=0.8\textwidth]{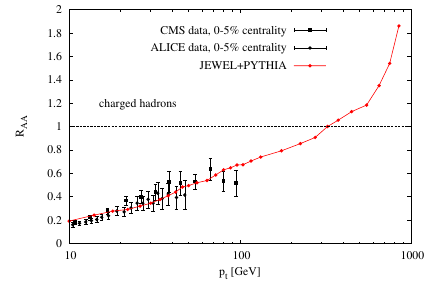}
\caption[Charged Hadron $R_{AA}$]{Hadron $R_{AA}$ as measured by CMS and ALICE collaborations compared to JEWEL predictions. Figure from \cite{zapp_perturbative_2013}.}
\label{charged_hadron_supression}
\end{figure}

\begin{figure}
\includegraphics[width=1.0\textwidth]{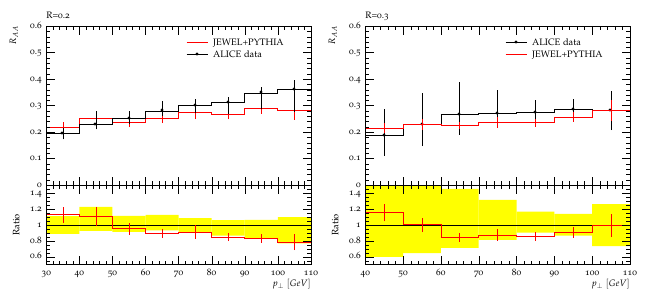}
\caption[JEWEL prediction of $R_{AA}$ for the jet $p_t$.]{Jet $p_T$ $R_{AA}$ as measured by CMS, ALICE and ATLAS collaborations compared to JEWEL predictions. Figure from \cite{zapp_perturbative_2013}.}
\label{jewel_jetpt_supression}
\end{figure}

JEWEL was developed to describe data from heavy-ion collisions. And it can reproduce most inclusive data. An example is displayed in Figure \ref{hadron_supression}. In the Figure we can see the prediction for neutral pion $p_t$ spectrum supression, as measured by PHENIX collaboration. It was one of the first results to indicate the phenomenum of Jet Quenching. In Figure \ref{charged_hadron_supression} we can see the supression for charged hadrons compared to data from ALICE and CMS collaborations. JEWEL describes this inclusive data really well. Also, in Figure \ref{jewel_jetpt_supression} we can see that the supression for reconstructed jets is also well described by JEWEL. The two Figures combined show that JEWEL can handle a wide energy range and different hadrochemistry well.

\begin{figure}
\includegraphics[width=1.0\textwidth]{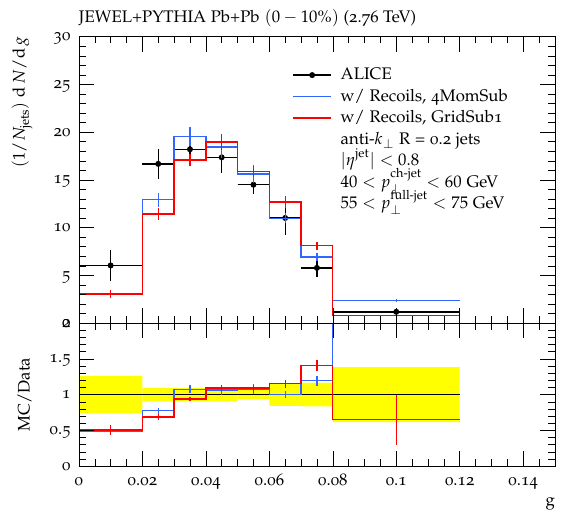}
\caption[JEWEL prediction for girth.]{JEWEL predictions for jet girth compared to measurements from ALICE collaboration. Figure from \cite{elayavalli_medium_2017}.}
\label{jewel_girth}
\end{figure}

\begin{figure}
\includegraphics[width=1.0\textwidth]{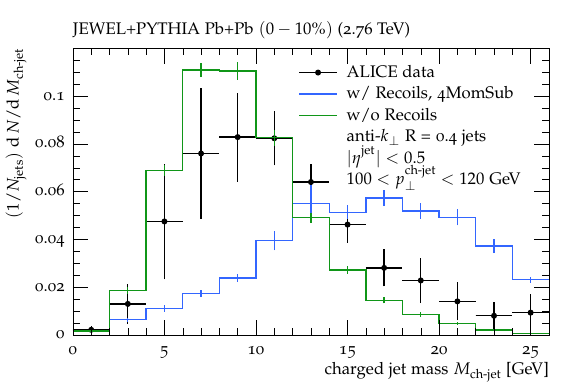}
\caption[JEWEL prediction for mass.]{JEWEL predictions for jet mass compared to measurements from ALICE collaboration. Figure from \cite{elayavalli_medium_2017}.}
\label{jewel_mass}
\end{figure}

Studying internal jet structure, we can see a somewhat different picture of JEWEL performance. For instance, in Figure \ref{jewel_girth} we see that JEWEL predicts jets broader than data. The jet mass can be seen in Figure \ref{jewel_mass}. We are interested in this work in the case with the recoiling scattering centers, since radiation patterns trying to probe the medium is our goal. JEWEL also predicts higher values for this observable, indicating broader jets than expected. In the case without recoils, the jet mass has lower values than data, which indicates a hardening of the core.

\mysection{JEWEL with realistic IC} \label{jewel_with_ic}

\begin{figure}
\includegraphics[width=1\textwidth]{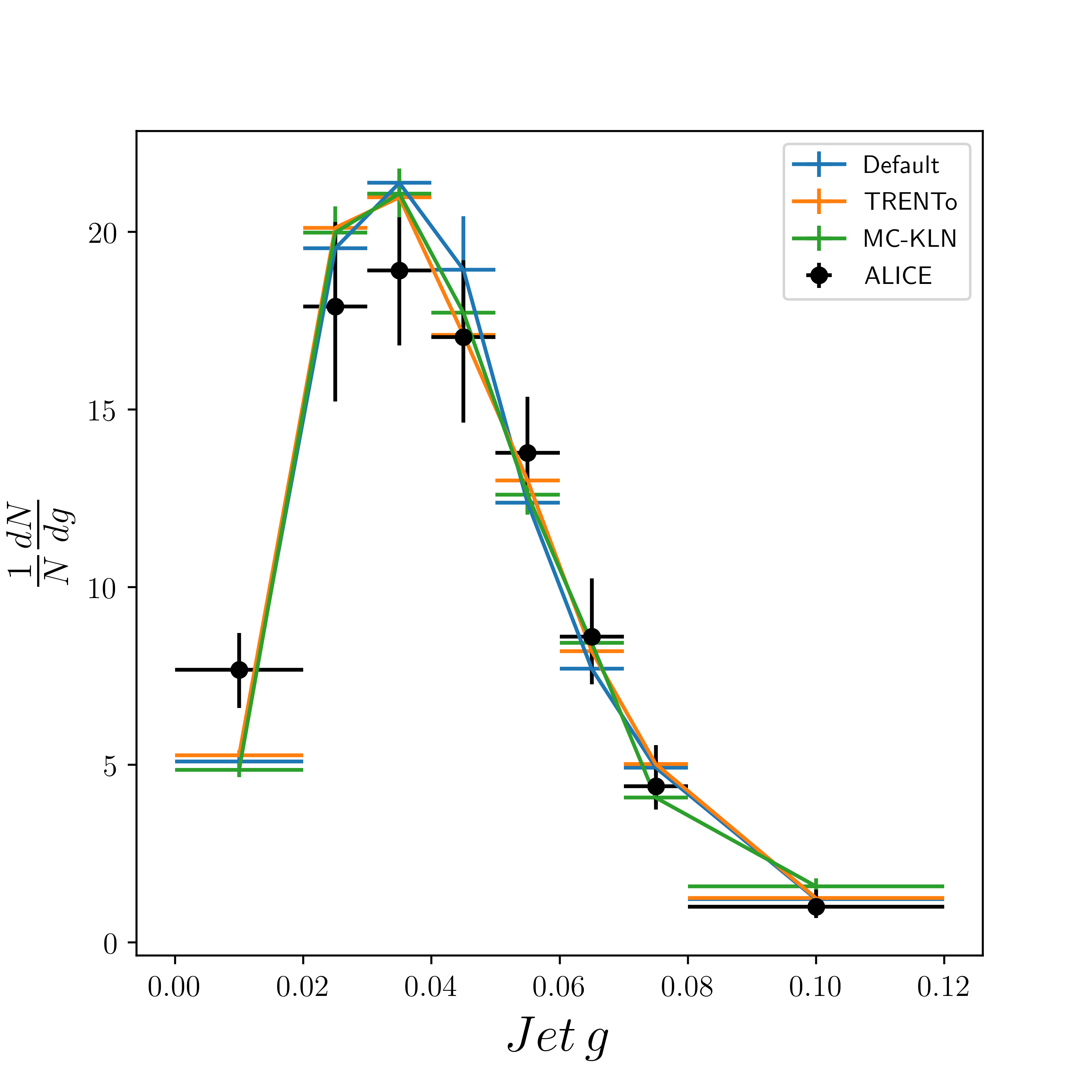} 
\caption[Jet Girth with realistic IC]{Jet Girth for charged jets calculated for $R=0.2$ anti-kt algorithm and $|\eta|<0.8$. $40 {\rm GeV/c} < p_T < 60 {\rm GeV}$. The CM energy is $\sqrt{s_{NN}}= 2.76 {\rm TeV}$. On the $0-10\%$ centrality class.}
\label{jet_girth_ic}
\end{figure}

In Figure \ref{jet_girth_ic} we see the results for jet girth with realistic initial conditions. Here $\rm T_RENTo$ was used with calibration to fit IP-Glasma results\cite{moreland_alternative_2015}. JEWEL tends to overestimate the peak value even with the new initial conditions. One is reminded that girth is related to the angular opening of the jet. Girth also depends linearly on the transverse momenta of the particles. The overestimation of JEWEL results tells us that the jets produced by the simulation tend to be slightly broader than the jets from data. No further improvement comes from the inclusion of more realistic initial conditions, as the results given by the inclusion of realistic initial conditions agree with the default of JEWEL.

\begin{figure}
\includegraphics[width=1\textwidth]{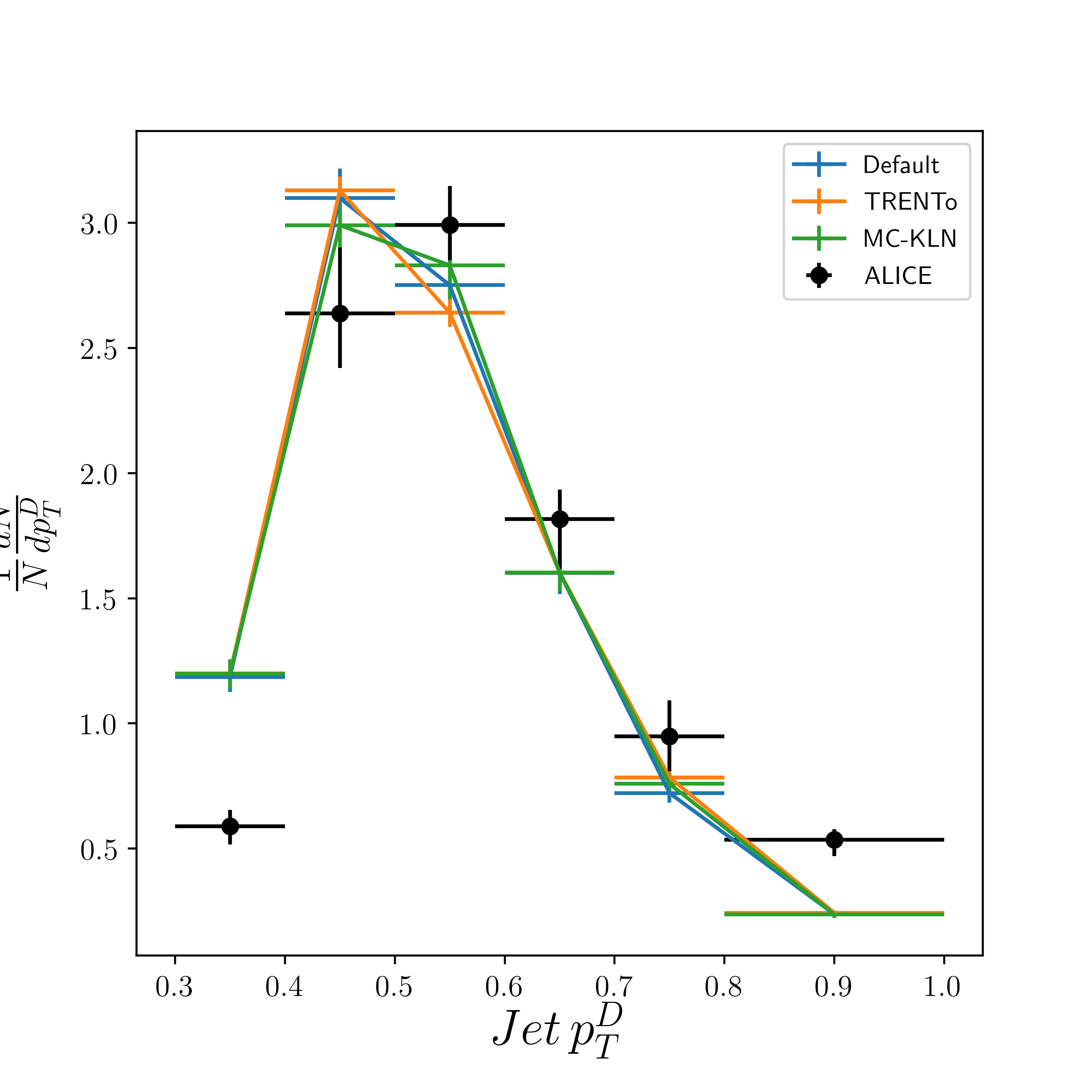} 
\caption[Jet $p_D^T$ with realistic IC]{Jet Dispersion for charged jets calculated for $R=0.2$ anti-kt algorithm and $|\eta|<0.8$. $40 {\rm GeV/c} < p_T < 60 {\rm GeV}$. The CM energy is $\sqrt{s_{NN}}= 2.76 {\rm TeV}$. On the $0-10\%$ centrality class.}
\label{jet_dispersion_ic}
\end{figure}

In Figure \ref{jet_dispersion_ic} we see the results for the jet dispersion. The default of JEWEl predicts slightly lower values than data. This indicates softer fragmentation. With the inclusion of realistic initial conditions, there is no substantial difference.

\begin{figure}
\includegraphics[width=1\textwidth]{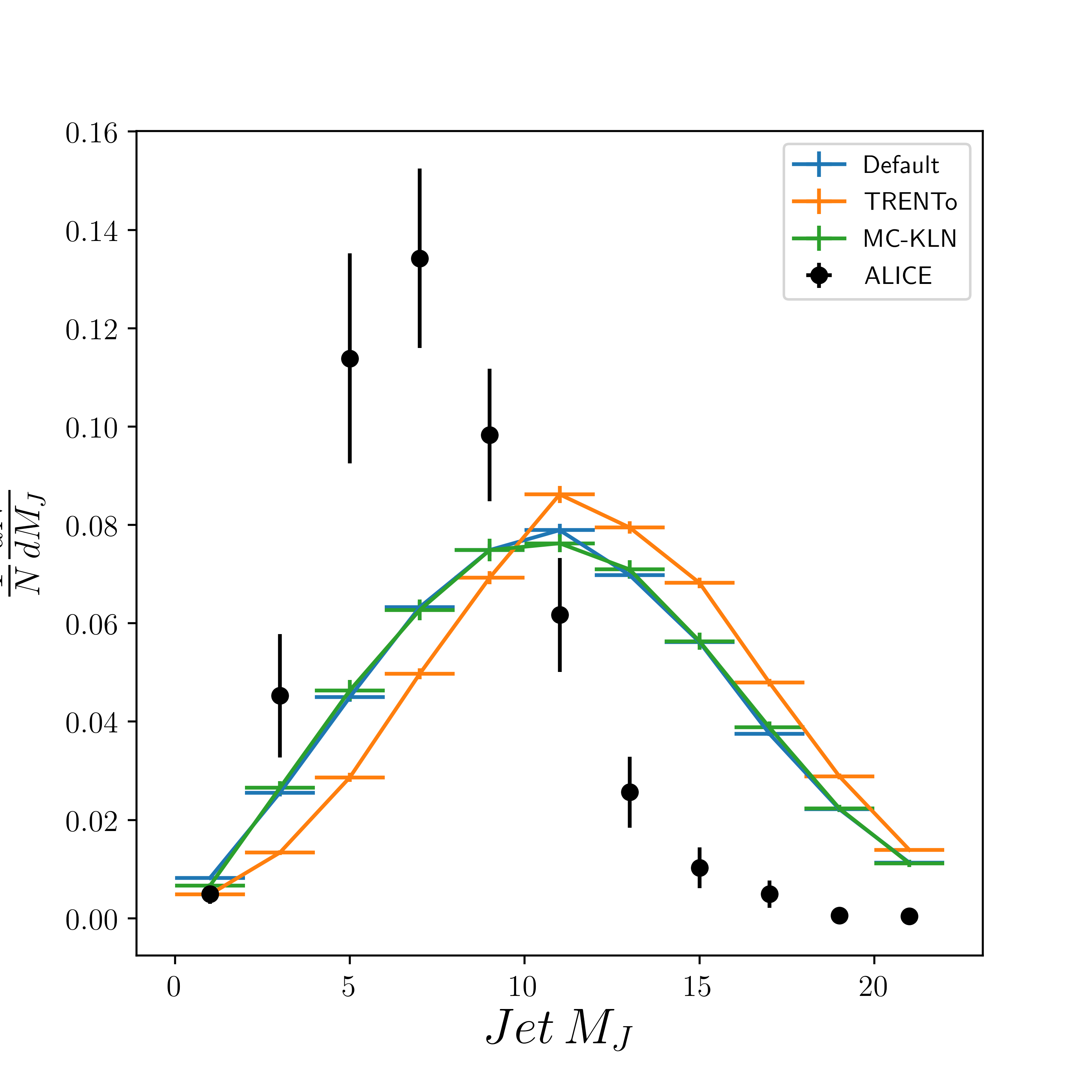} 
\caption[Jet Mass with realistic IC]{Jet Mass for charged jets calculated for $R=0.2$ anti-kt algorithm and $|\eta|<0.8$. $40 {\rm GeV/c} < p_T < 60 {\rm GeV}$. The CM energy is $\sqrt{s_{NN}}= 2.76 {\rm TeV}$. On the $0-10\%$ centrality class.}
\label{jet_mass_ic}
\end{figure}

The results for the jet mass are displayed in Figure \ref{jet_mass_ic}. Here we present also the inclusion of realistic IC for this observable. JEWEL in its default does not make a good prediction for it already. The problem is of the same nature of the disagreement of the girth, but worst. The jets in data tend to have a lower mass than predicted, this indicates larger jets, as is the case with girth. The mass further indicates that the problem lies in the soft fragmentation, which depends strongly on hadronization. There is some improvement from the addition of the realistic IC background, as one can see from the slight shift to the left. This shift is not significant though, due to the uncertainties of the method. There is a discrepancy in the spectrum. 

\begin{figure}\includegraphics[width=1\textwidth]{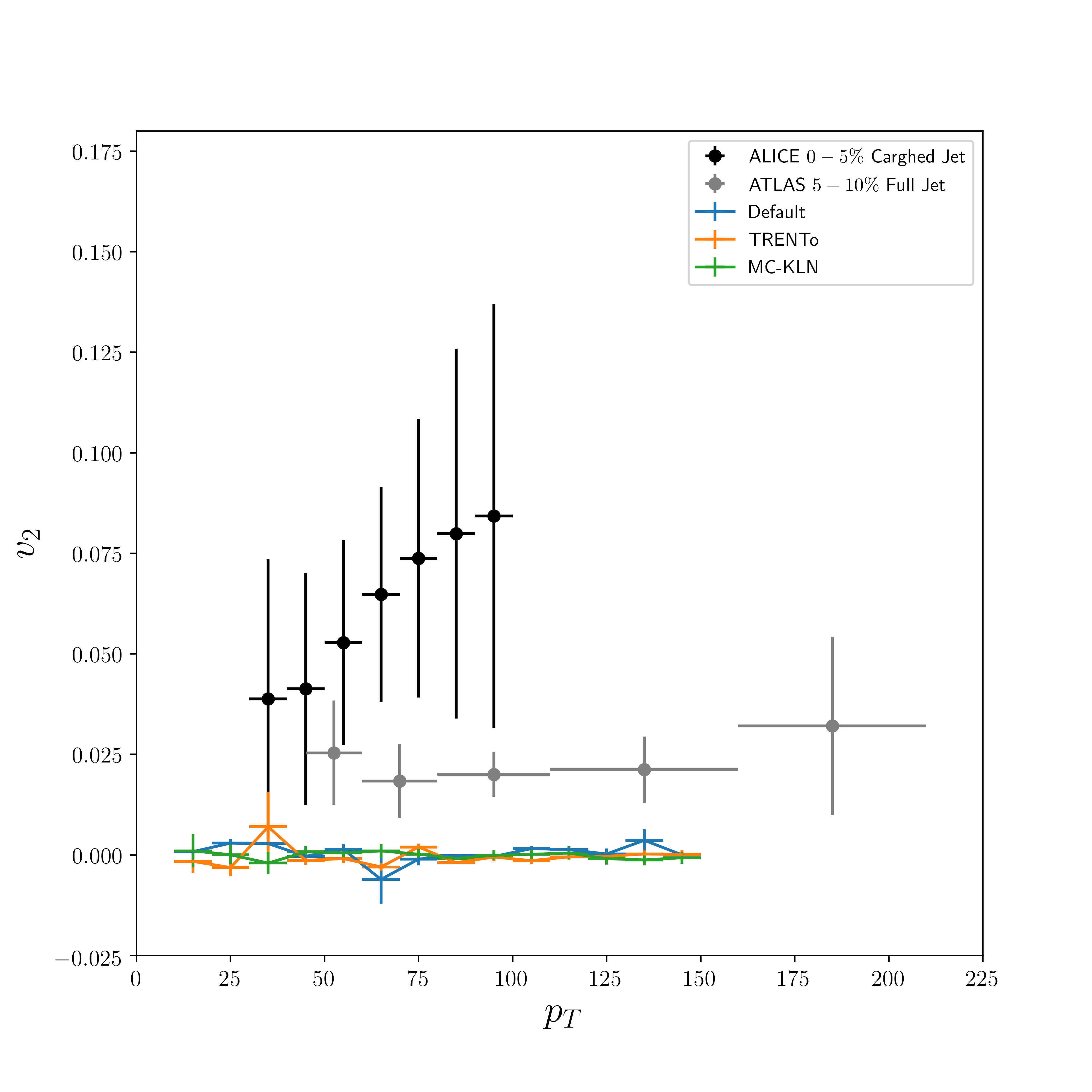} 
\caption[Jet $v_2$ with realistic IC]{Jet $v_2$ calculated for $R=0.4$ anti-kt algorithm and $|\eta|<0.8$. The CM energy is $\sqrt{s_{NN}}= 2.76 {\rm TeV}$. On the $0-10\%$ centrality class.}
\label{jet_v2_ic}
\end{figure}

In Figure \ref{jet_v2_ic} we see the results for the jet $v_2$. The data from ALICE and ATLAS show that there is tension between the experimental results making any conclusion about the performance of the model more difficult. ALICE uses their TPCs for jet reconstruction and ATLAS uses the hadronic calorimeters. This means that ALICE uses only charged particles, and ATLAS uses all hadrons for jet reconstruction. This explains why ALICE data has lower values of $p_T$. Although there is a disagreement between the collaborations, both data seems to indicate that the $v_2$ is different from zero. This is expected from fluctuations that happen on the initial conditions, since central collisions don't have a geometry that naturally raises an azimuthal asymmetry on the energy distribution. We can see in the Figure \ref{jet_v2_ic} that JEWEL, even with realistic IC, predicts values consistent with zero $v_2$.

\mysection{JEWEL with realistic hydro} \label{jewel_with_hydro}

\begin{figure}
\includegraphics[width=1\textwidth]{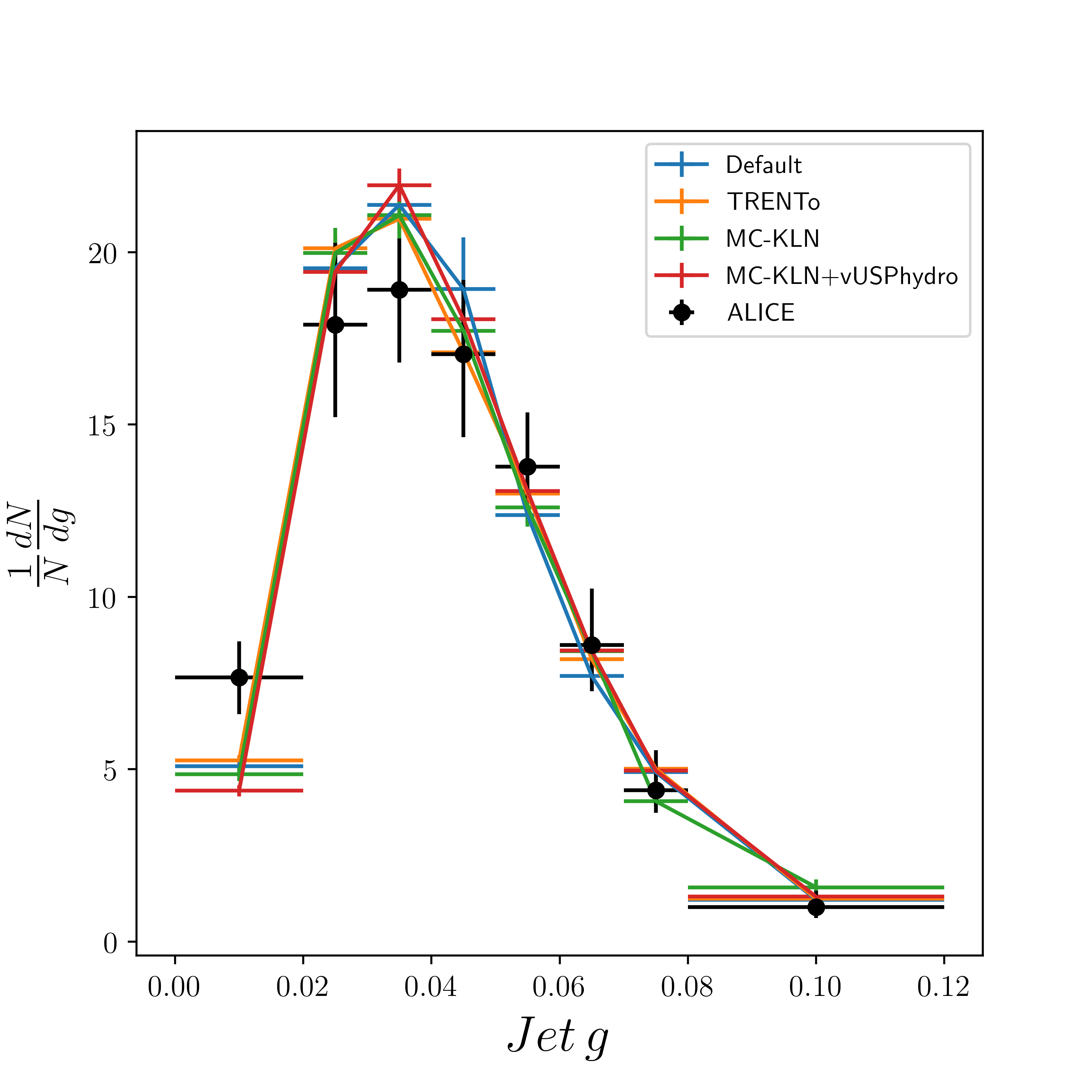} 
\caption[Jet Girth]{Jet Girth for charged jets calculated for $R=0.2$ anti-kt algorithm and $|\eta|<0.8$. $40 {\rm GeV/c} < p_T < 60 {\rm GeV}$. The CM energy is $\sqrt{s_{NN}}= 2.76 {\rm TeV}$. On the $0-10\%$ centrality class.}
\label{jet_girth}
\end{figure}

In Figure \ref{jet_girth} we see the results for jet girth. Considering the hydrodynamic expansion for the medium as well. The result shows that JEWEL tends to overestimate it. Since girth is related to the jet width, this shows broader jets than the data. No further improvement comes from the inclusion of more realistic initial conditions or hydrodynamics, as the results given by the inclusion of the realistic hydro and initial conditions agree with the default of JEWEL.

\begin{figure}
\includegraphics[width=1\textwidth]{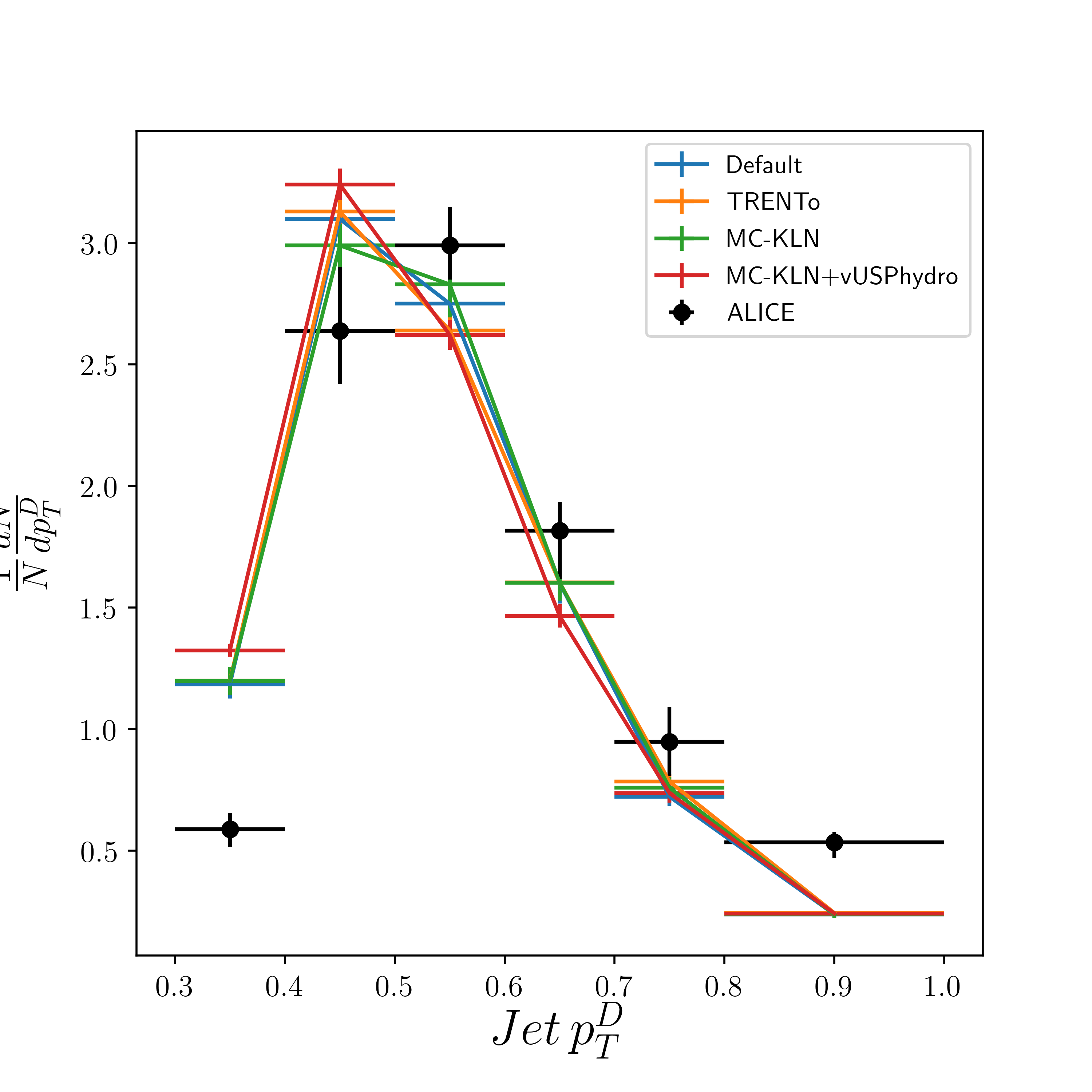} 
\caption[Jet $p_D^T$]{Jet Dispersion for charged jets calculated for $R=0.2$ anti-kt algorithm and $|\eta|<0.8$. $40 {\rm GeV/c} < p_T < 60 {\rm GeV}$. The CM energy is $\sqrt{s_{NN}}= 2.76 {\rm TeV}$. On the $0-10\%$ centrality class.}
\label{jet_dispersion}
\end{figure}

In Figure \ref{jet_dispersion} we see the results for the jet dispersion with the inclusion of realistic hydro. The default of JEWEl predicts slightly lower values than data. This indicates softer fragmentation. With the inclusion of realistic hydro, the agreement is slightly worst for lower values, but not substantially different. This indicates that the fragmentation is slightly softer than that of the default. Regions of greater density on the profile could be responsible for the further soft radiation causing this.

\begin{figure}
\includegraphics[width=1\textwidth]{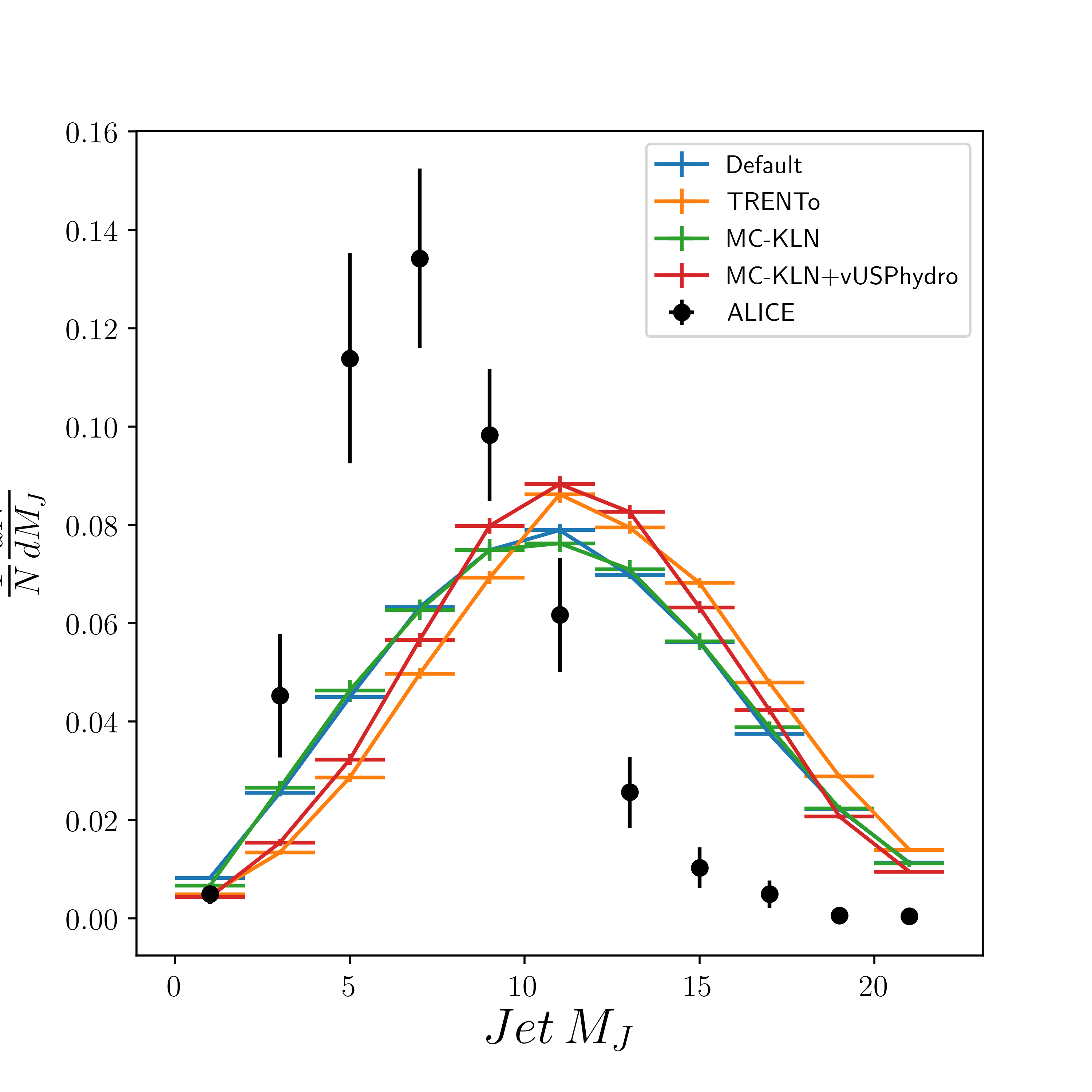} 
\caption[Jet Mass]{Jet Mass for charged jets calculated for $R=0.2$ anti-kt algorithm and $|\eta|<0.8$. $40 {\rm GeV/c} < p_T < 60 {\rm GeV}$. The CM energy is $\sqrt{s_{NN}}= 2.76 {\rm TeV}$. On the $0-10\%$ centrality class.}
\label{jet_mass}
\end{figure}

The results for the jet mass are displayed in Figure \ref{jet_mass}. JEWEL in its default doesn't make a good prediction for it. The problem is of the same nature of the disagreement of the girth, but worst. The jets in data tend to have a lower mass than predicted, this indicates larger jets, as is the case with girth. The mass further indicates that the problem lies in the soft fragmentation, which depends strongly on hadronization. There is some improvement from the addition of the realistic hydrodynamics background, although there is a discrepancy in the spectrum. 

In Figure \ref{jet_v2} we see the results for the jet $v_2$, with the inclusion of realistic hydro. Although there is a disagreement between the collaborations, both data seems to indicate that the $v_2$ is different from zero. The fluctuations might be responsible for the existence of the asymmetry. We can see in Figure \ref{jet_v2} that the inclusion of the realistic hydro has raised the $v_2$ from zero, although there seems to be overestimating it. The simulations currently performed tell us that the combined effect of realistic initial conditions and hydrodynamics are responsible for this effect. The fact that neither MC-KLN nor $\rm T_RENTo$ have shown significant $v_2$ indicates that the hydrodynamics is an essential ingredient for describing this observable.

\begin{figure}\includegraphics[width=1\textwidth]{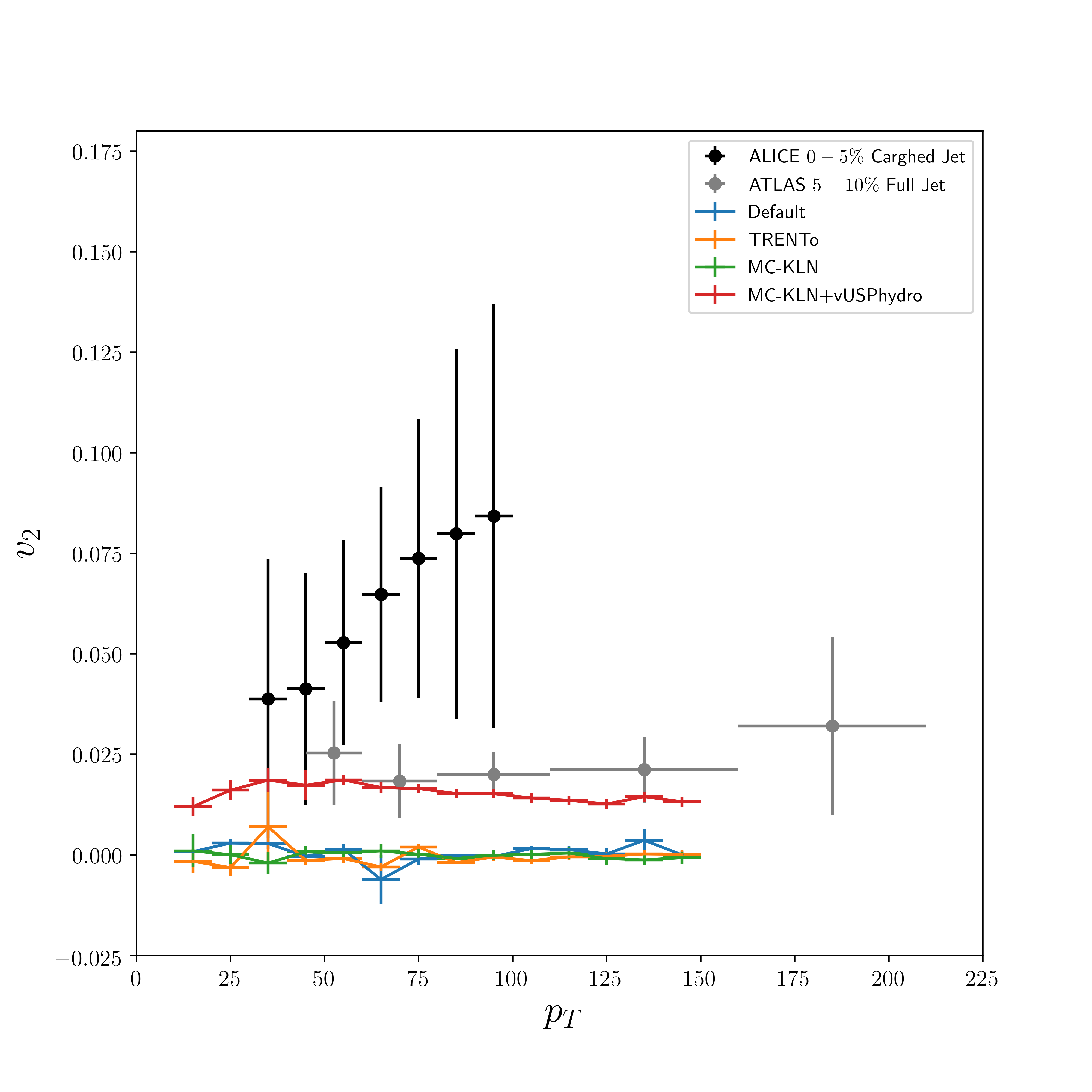} 
\caption[Jet $v_2$]{Jet $v_2$ calculated for $R=0.4$ anti-kt algorithm and $|\eta|<0.8$. The CM energy is $\sqrt{s_{NN}}= 2.76 {\rm TeV}$. On the $0-10\%$ centrality class.}
\label{jet_v2}
\end{figure}

\mychapter{Conclusion} \label{conclusions}

The results presented in the previous chapter show that, as far as the jet structure and shape are concerned, there are no major modifications of the observables due to the inclusion of realistic hydro and initial conditions. Some improvements came on the jet mass, bit only slightly.
\par
To improve the description of the data, other things could be modified. First, there is still the hadronization mechanism that is used by JEWEL, which comes from PYTHIA. This mechanism is built to explain pp data. Jets formed in heavy-ion collisions might be substantially different due to their hadronization mechanism. Effects of coalescence might play a big role here, giving new color partners to the leading partons describing the jets.
\par
Another thing that was not implemented in this work is the local four-velocity and a realistic EOS. This could further improve the description of the data. Since the partons are moving through the medium and the medium itself is most of the time moving outwards, the collisions could be more collinear than JEWEL default predicts. This could collimate the jet further, resulting in narrower jets. The collisions would be softer as well, since the CM of the local elastic collisions would be smaller, perhaps resulting in harder hadrons in the center of the jets in the final state. This could improve the $p_T^D$ data as well.
\par
The sample of the results presented in this work shows that there is not, as far as internal jet substructure and shape is concerned, a major improvement in describing the observables by implementing more realistic initial conditions or realistic hydrodynamics. Also, the correlation between medium and jets does present modifications due to the inclusion of a more realistic background, as indicated in the $v_2$ results. There is still high uncertainty in the experimental data, and further experimental improvement is needed for more detailed comparisons. The fact that this observable showed a value different from zero does indicate that high energy partons can be used to investigate properties of the medium through this kind of observable. Further information can be potentially extracted from the medium by looking at higher moments of the Fourier decomposition. This can aid in constraining the models applied in these predictions, as well as extract the transport coefficients from the QGP.

\bibliography{tese.bbl}
\markboth{Bibliography}{}
\bibliographystyle{ieeetr}

\printindex

\end{document}